\documentclass[12pt,preprint]{aastex}

\accepted{}
\journalid{}{}
\articleid{}{}

\lefthead{Mirabal {\it et al.}}
\righthead{Time-Dependent Optical Spectroscopy of GRB 010222}

\received{2002 March 4}
\begin{document}

\def\etal{{\it et al.}}
\def\eg{{\it e.g.,}}
\def\ie{{\it i.e.,}}
\def\vs{{\it vs.}}
\def\etc{{\it etc.}}
\def\kms{km~s$^{-1}$}
\def\Msol{M$_\odot$}
\def\lsim{\mathrel{\lower .85ex\hbox{\rlap{$\sim$}\raise
.95ex\hbox{$<$} }}}
\def\gsim{\mathrel{\lower .80ex\hbox{\rlap{$\sim$}\raise
.90ex\hbox{$>$} }}}

\def\pz{\phantom{0}}
\def\lsim{\mathrel{\lower .85ex\hbox{\rlap{$\sim$}\raise
.95ex\hbox{$<$} }}}
\def\gsim{\mathrel{\lower .80ex\hbox{\rlap{$\sim$}\raise
.90ex\hbox{$>$} }}}
\def\bv{($B-V$)}
\def\vr{($V-R$)}
\def\br{($B-R$)}
\def\ub{($U-B$)}
\def\vi{($V-I$)}
\def\ri{($R-I$)}
\def\source{3EG~J1835+5918}
\def\ro{{\it ROSAT\/}}
\def\asca{{\it ASCA\/}}

\title{Time-Dependent Optical Spectroscopy of GRB 010222: Clues to the GRB 
Environment\altaffilmark{1}}

\author{N. Mirabal\altaffilmark{2}, J. P. Halpern\altaffilmark{2},
S. R. Kulkarni\altaffilmark{3}, S. Castro\altaffilmark{3,4},
J. S. Bloom\altaffilmark{3}, S. G. Djorgovski\altaffilmark{3},
T. J. Galama\altaffilmark{3},  F. A. Harrison\altaffilmark{3},
D. A. Frail\altaffilmark{6}, P. A. Price\altaffilmark{3}, 
D. E. Reichart\altaffilmark{3} and
H. Ebeling\altaffilmark{5}
}
\altaffiltext{1}{Based on data obtained at the W.M. Keck Observatory,
which is operated as a scientific partnership among the California 
Institute of Technology, the University of California, and NASA, and
was made possible with the generous financial support of the W.M. Keck 
Foundation}
\altaffiltext{2}{Astronomy Department, Columbia University, 550 West 120th 
Street, New York, NY 10027}
\altaffiltext{3}{California Institute of Technology, Palomar Observatory 
105-24, Pasadena, CA 91125}
\altaffiltext{4}{Infrared Processing and Analysis Center, 100-22
California Institute of Technology, 105-24, Pasadena, CA 91125}
\altaffiltext{5}{Institute for Astronomy, University of Hawaii, 2680 
Woodlawn Drive, Honolulu, HI 96822}
\altaffiltext{6}{National Radio Astronomy Observatory, Socorro, NM 
87801}

\begin{abstract}
\rightskip 0pt \pretolerance=100 \noindent
We present sequential optical spectra of the afterglow of 
GRB 010222 obtained one day apart using the Low Resolution
Imaging Spectrometer (LRIS) and the Echellette Spectrograph and Imager
(ESI) on the Keck telescopes. 
Three low-ionization absorption systems are spectroscopically identified 
at $z_{1}=1.47688$, $z_{2}=1.15628$, and $z_{3}=0.92747$.
The higher resolution ESI spectrum reveals
two distinct components in the highest redshift system at
$z_{1a}=1.47590$ and  $z_{1b}=1.47688$.
We interpret the $z_{1b}=1.47688$ system as an absorption feature
of the disk of the host galaxy of GRB 010222. 
The best fitted power-law optical 
continuum and [Zn/Cr] ratio imply low dust content 
or a local gray dust component near the burst site.
In addition, we do not detect strong signatures of
vibrationally excited states of $H_{2}$.
If the GRB took place in 
a superbubble or young stellar cluster,
there are no outstanding signatures of an ionized absorber, either.
Analysis of the spectral time dependence at low resolution 
shows no significant evidence for absorption-line variability.
This lack of variability is
confronted with time-dependent photoionization simulations 
designed to apply 
the observed flux from GRB 010222 to a variety of assumed atomic gas 
densities and cloud radii.  
The absence of time dependence in the 
absorption lines implies that 
high-density environments are disfavored.
In particular, if the GRB environment was dust free, its density was unlikely
to exceed $n_{H}=$$10^{2}$ cm$^{-3}$. If depletion of metals onto dust 
is similar
to Galactic values or less than solar abundances are present, then $n_{H}$ 
$\geq$ 2 $\times$ $10^{4}$ cm$^{-3}$
is probably ruled out in the immediate vicinity of the burst.
   
\end{abstract}


\section{Introduction}

Years after the serendipitous
discovery of gamma-ray bursts (GRBs) by the Vela 
spacecraft (Klebesadel et al. 1973), the nature 
of the progenitors responsible for generating 
the initial explosion remains uncertain, while
an elegant set of ideas about the production and evolution of the 
afterglow has developed (M\'esz\'aros \& Rees 1997; Sari \& Piran 1997). 

Aside from recent breakthroughs in X-ray 
spectroscopy of GRBs (\eg\ Piro et al. 2000)
a large fraction of the observational information about 
the environments of GRBs derives  
from optical spectroscopy and imaging of the host galaxies. Optical 
spectroscopy of the integrated light and calibrated emission lines 
has been used 
to derive the star-formation rate (SFR) for a number of the host 
galaxies (see for instance Bloom et al. 1998; 
Djorgovski et al. 1998). As a complement to spectroscopy, 
high-resolution optical imaging and astrometry have helped 
pinpoint two-dimensional locations for GRBs with respect to 
the host galaxy. The distribution of these GRB locations 
has yielded a number of statistical constraints 
in the progenitor scenarios (\eg\ Bloom, Kulkarni, \& Djorgovski 2000). 

A promising technique that has been applied with less success 
so far is the study of time dependence in absorption lines
from metal ions (Perna \& Loeb 1998; B\"ottcher et al. 1999)
or ${\rm H}_2$ vibrational levels (Draine 2000) that might be excited
by the UV radiation generated during the evolution of the burst. 
The detection of absorption-line variability could provide important clues 
about the physical dimensions of the photoionized region 
and the density in the immediate environment of the GRB.
Absorption-line variability can be 
quantified by measuring changes in the equivalent 
widths  as a function of time or by identifying the 
appearance of new absorption features. 
Vreeswijk~et al. (2001) studied the time evolution of 
the Mg II doublet in the optical spectra
of GRB 990510 and GRB 990712 but failed to find any significant changes.
In this work we present a study of the time evolution of absorption systems
in the optical spectrum of GRB 010222, and discuss possible implications
for its progenitor environment. The outline of the paper
is as follows: \S 2 describes the optical spectroscopy,   \S 3 describes
the absorption line identification and continuum fitting. In \S 4 we detail
column density determinations, arguments in favor of the
$z=1.47688$ redshift for GRB 010222, kinematics of the host galaxy, and
a study of time evolution of absorption lines.  A description of 
the photoionization
code and results is given in \S 5 and \S 6. Finally,  
the implications of our results and conclusions are 
presented in \S 7 and \S 8.

\section{Observations}

\subsection{LRIS Spectroscopy}

GRB 010222 was initially detected by BeppoSAX on UT 2001 
Feb. 22.30799 (Piro 2001). 
An optical counterpart was reported only a few hours later by Henden 
(2001a,2001b). Our group began 
spectroscopic observations at the position of the optical counterpart 
on UT 2001 February 22.66 
with the dual-beam Low Resolution Imaging Spectrometer (LRIS) on 
the Keck I telescope and obtained a second spectrum on UT 2001 
February 23.66.  On both nights we used a 300 lines/mm 
grating blazed at 5000 \AA\ with a $1^{\prime\prime}\!.5$ wide
slit. The effective spectral resolution FWHM varies 
from $\approx$ 13 \AA~on the blue 
side to  $\approx$ 11 \AA~on the red side. 
A total of 1800~s of exposure were 
obtained on  each night. The data were
trimmed, bias-substracted and flat-fielded using standard procedures.
Extraction of the spectra was performed using the IRAF APALL package.
Telluric lines were corrected by fitting the continuum including 
the A and B atmospheric bands of a spectrophotometric
standard. The spectra of spectrophotometric 
standards Feige 67, Hiltner 600, and HZ 44 were used for flux calibration and a
Hg-Kr-Ne-Ar-Xe comparison lamp provided the wavelength calibration. 
The wavelength calibrations have a typical accuracy of
$\approx 0.2-0.5$ \AA\ as determined by the root-mean-square (rms) deviation of
the individual lines from the smooth function used to fit them.
The large range of precision reflects the presence of fewer arc lines
on the blue side.  Finally, the LRIS-B (blue channel) and LRIS-R 
(red channel) spectra were connected at the crossover wavelength
of the dichroic beam splitter.

\subsection{ESI Spectroscopy}

A higher-resolution spectrum of GRB 010222 was obtained on UT 2001 Feb 23.61
using the Echellette Spectrograph and Imager (ESI) mounted
on the Keck II telescope. 
This mode spans ten orders with effective 
wavelength coverage from 3900 \AA~to 10900 \AA. The spectral 
resolution is 11.4 km s$^{-1}$ pix$^{-1}$ with typical 
dispersion in the instrument ranging from 0.16 \AA~pixel$^{-1}$
in order 15 to 0.30 \AA~pixel$^{-1}$ in order 6. 
In the echellette mode the instrument provides a wavelength resolution  
R $\approx$ 10000 over the entire spectrum.
A total of 3 $\times$~1200 second exposures
were obtained and reduced using Tom Barlow's 
Mauna Kea Echelle Extraction program (MAKEE)\footnote{See http://spider.ipac.caltech.edu/staff/tab/makee/}. MAKEE, originally designed
for the High Resolution Echelle Spectrometer (HIRES), is an   
automated procedure that allows 
standard processing as well as wavelength and flux normalization.
For the wavelength calibration we used a CuAr cathode lamp which 
provides a large number of unsaturated lines to fit the spectrum with 
rms $\approx$ 0.09 \AA. The individual spectra at a common epoch were  
average-combined after running through the pipeline extraction.

\section{Reductions}

\subsection{Absorption System Identifications}

We have identified three absorption 
systems (Bloom et al. 2001; Castro et al. 2001a) 
which were reported independently by Jha et al. (2001).
The systems are located at $z_{1}=1.47688$, $z_{2}=1.15628$, and 
$z_{3}=0.92747$, where the redshift reported is the average of individual 
lines for each system. The  line identification 
corresponds to a cross-correlation of features in the spectrum with
a listing of known absorption lines (Verner, 
Barthel, \& Tytler 1994) assuming possible redshift
systems at $z_{1}$, $z_{2}$, and $z_{3}$. In order to weed out spurious 
features possibly due to sky subtraction, 
only those lines present in three spectra are reported. Moreover, we demanded
$\lambda_{rest}$[1.0 + z(1-$x$)] $<$  $\lambda_{obs}$ $<$  
$\lambda_{rest}$[1.0 + z(1+$x$)] and the corresponding 
oscillator strength ratios for identified lines. In the previous,
$x$~stands for a tolerance factor 
in the line identification and accounts for
uncertainties in the wavelength calibration. 
Figure 1 shows the LRIS spectra including 
line identification for each absorption system. The 
higher-resolution ESI spectrum with identified 
absorption features is shown in Figure 2. 
In the highest redshift system we find kinematic evidence
for two distinct components at  
$z_{1a}$=1.47590 and  $z_{1b}$=1.47688 respectively, 
with a rest-frame separation of $\approx 119$~km~s$^{-1}$.
The kinematic structure
of this system will be discussed at greater length in \S 4.3.
Tables 1 and 2 summarize the line identifications for each setup and 
list the measured vacuum wavelengths, 
observed wavelengths converted into vaccum and 
corrected for heliocentric motion, estimated redshift,
oscillator strengths $f_{ij}$, 
equivalent widths ($W_{o}$) in the rest frame for both nights, and 
error estimates in their equivalent widths. 
We have adopted the compilation of wavelengths and oscillator 
strengths by Morton (1991)
with the revisions proposed by Savage \& Sembach (1996).
The main source of uncertainty in the equivalent width
measurements derives from the identification of the continuum level. 
In order to compute the errors in 
the equivalent width for each line we used the 
IRAF SPLOT task which allows error estimates based on a Poisson 
model for the noise. 
For blended lines, IRAF SPLOT fits and deblends each line 
separately using predetermined line profiles.  
Error estimates for blended lines are computed directly 
in SPLOT by running a number of Monte-Carlo simulations based on  
preset instrumental parameters.

\subsection{Continuum Fitting}

The flux-calibrated spectra were corrected for Galactic foreground 
extinction assuming $E(B-V) = 0.023$ (Schlegel, Finkbeiner, \& Davis 1998)
and the extinction curve of Cardelli, Clayton \& Mathis (1989). 
After correcting for Galactic extinction, the spectra were
converted to $F_{\nu}$ and $\nu$ units and 
the continuum divided in equal size bins, excluding regions containing 
absorption lines. The statistical 1$\sigma$ error 
was determined by estimating the standard deviation within each bin. 
Next we found that a
power law of the form $F_{\nu} \propto \nu^{\beta}$ produced an adequate 
fit to the optical spectrum. 
The best fitted model changed slightly from 
$\beta = -0.89 \pm 0.03$ on 
Feb. 22.66 to $\beta = -1.02 \pm 0.08$ on Feb. 23.66.
The continuum points along with the best fitted models are shown in 
Figure 3. Despite the apparent steepening of the spectra, 
there is one important caveat, the difficulty
of doing spectrophotometry through a narrow slit. Because we have 
used a 1.5'' slit width the throughput of the spectrograph is reduced. 
Spectrophotometry might also 
be affected  by atmospheric differential refraction (Filippenko 1982). 
On Feb 22.66 this effect was minimized by aligning the slit  
at the parallactic angle
for the GRB as well as for the spectrophotometric 
standards. Moreover all the objects were observed at similar airmasses. 
During Feb 23.66 only one slit angle was used throughout the night and
the airmasses for the spectrophotometric standards were slightly higher at
the time of acquisition. Therefore, 
the uncertainty in the spectrophotometry during the second night  
still allows the possibility that the spectral index did not change 
between the two nights.
The difficulty of achieving accurate spectrophotometry in this case 
gives more weight to the results obtained from 
simultaneous multi-band photometry. Our derived spectra index
$\beta = -0.89 \pm 0.03$ appears 
in agreement with the reported single power law $\beta = -0.88 \pm 0.03$
obtained from {\it BVRI} optical data (Stanek et al. 2001).  Moreover 
the spectral behavior of the lightcurves does not show any steepening of
the index $\beta$ as a function of time (Stanek et al. 2001).  
The previous result 
is also consistent with $\beta = -0.89 \pm 0.03$
derived from a low-resolution spectrum obtained
five hours after the burst (Jha et al. 2001). But our spectral index is 
slightly shallower than  the index $\beta = -1.1 \pm 0.1$ 
reported by Masetti et al. (2001). One possible cause for the steepening
in the Masetti et al. (2001) results
is the inclusion of 
the {\it U}~band data in the overall fit (Stanek et al. 2001).
The bulk of the evidence seems to argue for a single power law spectral index 
for the duration of the burst.  Hereafter we adopt $\beta = -0.89 \pm 0.03$. 

Our optical spectral index is similar 
within uncertainties to the measured 
X-ray spectral index $-0.97 \pm$ 0.05 reported for the whole BeppoSAX 
observation (in't Zand et al. 2001). However, as pointed out by 
these authors, the optical flux is too faint or the X-ray flux is
too bright to accommodate a simple synchrotron power-law.
One possible explanation 
for this mismatch is an X-ray excess above the standard 
synchrotron spectrum due 
to inverse Compton (IC) 
scattering (Sari \& Esin 2001). Harrison et al. (2001) 
have recently presented broad-band observations of
GRB 000926 that are consistent with this IC interpretation.  
An alternative explanation for the difference between the optical and X-ray 
flux is reddening of the optical spectrum. 
The steep slope of the continuum fit to the optical spectra 
in addition to the absence
of a 2175 \AA~``bump'' ubiquitous in a Galactic extinction curve 
implies that the extinction curve intrinsic to GRB 010222
would have to be approximately flat in order to not redden the spectra 
dramatically.
This in turn suggests that the dust content at the burst location is 
relatively
low or that any dust present mimics the gray variety (Aguirre 1999).  
IR and submillimeter observations of GRB 010222 by Frail 
et al. (2001b) concluded that the majority of the reprocessed emission
seen at the host galaxy is from a separate starburst component. By their 
estimation thermal radiation or dust scattering
near the burst site cannot reproduce the bulk of
the observed sub-millimeter energy.
Considering the possibility that the host is a starburst galaxy, we
estimated the intrinsic reddening 
using the extinction curve of Calzetti et al. (2000). Dust corrections are
particularly dangerous 
given the complexity of the dust geometry in any particular
line of sight (Witt et al. 1992). Nonetheless, even
a small color excess, $E(B-V) \approx 0.1$, requires a flatter intrinsic power
law for the afterglow with $\beta = -0.39$. This color excess is 
required to satisfy
$\beta = -(p-1)/2$ for $\nu < \nu_c$ in the optical band and $\beta = -p/2$ at 
frequencies $\nu > \nu_c$ in the X-ray regime.  Here $p$ is the index
of the power-law electron energy distribution, and $\nu_c$ is the
``cooling frequency'' at which the electron energy loss time scale
is equal to the age of the shock.  The spectrum,
corrected for such extinction at the host galaxy, is illustrated in Figure 4.
Alternatively, it is possible that  
$\nu_c$ has moved below the optical band at the time of our first 
spectrum, 8.45 hr after the burst.  If this is the case, the reddening
argument does not apply; instead, an IC scattering interpretation
is possible.

\section{Analysis}

\subsection{Metallic Absorption Systems}

The statistical properties of metallic absorption systems have 
been studied extensively at various spectral 
resolutions (\eg\ Steidel \& Sargent 1992;
Churchill et al. 1999). These studies have provided insight into
the equivalent width distribution, evolution, and cosmological clustering of 
absorbers. In the particular case of GRB 010222, the equivalent 
widths 
of the Mg II doublet in the identified systems 
indicate that they are unusually strong in comparison to the population
of absorbers studied by Steidel \& Sargent (1992). 
The Mg II doublet ratios 
$W_{o}$(Mg II 2796.35)/$W_{o}$(Mg II 2803.53) 
are 1.06 $\pm$ 0.06,
1.17 $\pm$ 0.19, and 1.46 $\pm$ 0.12 for the $z_{1}$=1.47688,  
$z_{2}$=1.15628 and
$z_{3}$=0.92747 systems, respectively. In each case the ratio differing from 
2 implies that the Mg II lines are strongly saturated. In addition to the 
saturation of Mg II lines, we shall show evidence for saturation effects in 
the majority of Fe II lines in the ESI spectrum.
We avoided using these lines for direct column density measurements
given the likelihood of a non-standard curve of growth
under saturated conditions (\eg\ Morton \& Bhavsar 1979).  

Instead, resolved absorption lines in the 
ESI spectrum that are not strongly
saturated are more appropriate
for column density determinations. The  
Mn II triplet at 2576.88 \AA, 2594.50 \AA, and 2606.46 
\AA\ in the $z_{1}$=1.47688 system 
meets this requirement and appears particularly well suited to
studying the highest column absorption system at
$z_{1b}$ in which it resides. These lines 
fall in the region of intermediate optical 
depth, $\tau \geq 1$, {\it i.e.}, on the ``flat'' part
of the curve of growth. Here the observed
broadening of the Mn II lines is assumed to be due to 
the random motion of the gas, described by a Maxwellian velocity
distribution. In this regime the equivalent width 
becomes sensitive to the Doppler parameter $b$, which can be determined 
directly from the broadening of the absorption lines. The
measured Doppler parameters $b$ are listed in Table 3.
In order to derive column density $N_{j}$ for any metal $j$,
we used the general prescription
for a standard curve of growth (Spitzer 1978) 
	
\begin{equation}
\frac{W_{\lambda}}{\lambda} = \frac{2bF(\tau)}{c}
\end{equation}

\begin{equation}
\tau = \frac{1.497 \times 10^{-2} N_{j} \lambda f_{ij}}{b}
\end{equation}

\noindent
where $N_{j}$ is written as a function of 
equivalent width $W_{\lambda}$, rest wavelength $\lambda$, oscillator 
strength $f_{ij}$, optical depth $\tau$, and
its dependence on the Maxwellian velocity 
distribution F($\tau$). Metallic 
column densities log($N_{j}$) as well as equivalent
neutral hydrogen column densities log($N_{\rm HI}$) 
assuming solar abundances (Anders \& Grevesse 1989) 
are given in Table 3. 
Notice the overall agreement in the column densities derived 
using three different Mn II lines. 
The column densities listed in the fifth and sixth columns
of Table 3 have not been corrected
for depletion into dust grains (Jenkins 1987), which is an
uncertain quantity.  

Waxman \& Draine (2000) have 
shown that dust may be sublimated by a prompt optical/UV flash at any distance
closer than $\approx$ 10 pc from the burst. Fruchter, Krolik 
\& Rhoads (2001) also  
considered the possibility of grain charging as a mechanism for dust 
destruction. This could take place as far away as $\approx$ 100 pc from
the birthsite. Thus, the combined effect of both destruction mechanisms, 
temperature of the medium,
and the return of dust depleted metals to the gas 
phase make the depletion correction largely uncertain. For completeness 
we make the simplest assumption ignoring any immediate return of metals into 
the gas phase and adopt a depletion correction of 
$-1.45$ dex for Mn$^{+}$ based 
on Goddard High-Resolution Spectrograph (GHRS) measurements towards
$\zeta$ Oph obtained by Savage \& Sembach (1996). 
The resulting
corrected values are shown in the last column of Table 3. Given the 
uncertainty in the amount of depletion it appears safe to regard the 
uncorrected HI column densities as lower limits. We note that the  
weighted average of Mn$^{+}$ corrected for depletion implies
log$(N_{\rm HI})_{\rm corr} = 21.89 \pm 0.13$ cm$^{-2}$ in the $z_{1a}$
system, which is similar to the 
redshift-corrected column density log$(N_{\rm HI}) \approx 22.00 $ cm$^{-2}$
derived by in 't Zand et al. (2001) from the X-ray afterglow. 
The results presented so far assume solar abundances, alternatively, 
less than solar abundances for Mn II 
might accommodate the difference between metal
abundances and the values reported by in 't Zand et al. (2001) and 
Salamanca et al. (2001).

We have completed the estimates of column densities $N_{j}$ by including 
blended absorption lines. The term ``blended'' is used here to refer to 
the lines where the $z_{1a}$ and $z_{1b}$
systems cannot be separated cleanly in the ESI spectrum. In
particular, by comparing the equivalent widths and the 
oscillator strength ratios for 
Zn II (2026.14 \AA, 2062.66 \AA) and Cr II (2056.25 \AA, 2062.23 \AA, 2066.16 \AA)
lines we see that they fall on the linear portion of the curve of growth.
For these lines we can obtain a direct measurement of the column density. On
the other hand, the Fe$^{+}$ and Mg$^{+}$ lines, although structured, are 
strongly saturated.  For that reason we decided to fit Gaussian profiles 
at the positions of the identified $z_{1a}$ and $z_{1b}$ components
while running the deblending routines. In this case the 
linear part of the curve of growth only provides a lower limit 
to the column density $N_{j}$ through the equation

\begin{equation}
N_{j}({\rm cm}^{-2})= 1.13 \times 10^{17}\frac{W_{\lambda}({\rm m\AA})}{f_{ij} 
\lambda^{2}({\rm \AA})}
\end{equation}

\noindent
The resulting column densities are listed in Tables 3 and 4.
In each case the density corresponds to the average of various 
single transitions. As a consistency check,
the HI column density derived assuming solar abundances for 
the generally undepleted Zn$^{+}$ is in agreement with the
values cited in Table 3 for the resolved Mn$^{+}$ lines. 

\subsection{Is $z_{1}$=1.47688 the redshift of GRB 010222?}

The absence of any emission lines in the spectrum
slightly complicates 
the unequivocal identification of this system with the host galaxy. 
However, there is information in the optical spectrum that supports
this interpretation. First, the non-detection
of a Lyman-$\alpha$ break places 
an upper limit for the GRB host galaxy of z $\leq$ 1.96. Second, Salamanca
et al. (2001) have reported evidence of a red wing of Ly-$\alpha$
absorption at $z \approx$ 1.476.  Third,
the measured equivalent widths of the Mg II doublet and the ratio of
Mg I to Mg II in this system are also among
the highest known for GRB afterglows and for the sample of
Mg II absorbers compiled 
by Steidel \& Sargent (1992) at similar spectral resolution.
Fourth, we can quantify the 
probability of detecting three intervening galaxies in the line of 
sight.  The LRIS setup used during the first two nights
spans the 3600--9205 \AA~wavelength range with 
a rest equivalent width threshold limit $W_{o} \geq$ 0.3 \AA, which
matches the resolution and equivalent width detection
limit of the Steidel \& Sargent survey. 
This wavelength coverage can be translated into 
a redshift range for Mg II lines of $0.284 \leq z \leq 2.284$,
or a redshift  interval $\Delta z = 2.0$. Thus, the number of 
absorbers per unit redshift is N/$\Delta z = 1.5$,
which is less stringent than an earlier estimate of this 
quantity (Jha et al. 2001), but is still larger than the expected number 
of Mg II absorbers per unit redshift 
$\langle N/z \rangle$ = 0.97 $\pm$ 0.10 for a rest equivalent width
threshold of 0.3 \AA~(Steidel \& Sargent 1992). The probability for the  
chance superposition of three absorbers in this line of sight is non-negligible
but small.
Lastly, if the highest redshift of an MgII system is $z$=1.47688,
    then, in 80\% of the cases, the source could have originated from
    no further than $z\approx 1.8$. If it originated from a larger
    distance then, given the rapid increase in the number
    of metal absorption systems with increasing redshift, a source
    redshift higher than z=1.8 would be unlikely (Bloom et al. 1997).
Although not definitive proof, these arguments
are most consistent with the $z_{1}$=1.47688 absorption
system as the host galaxy of GRB 010222. 

\subsection{Kinematics and abundances of the $z_{1}$=1.47688 absorption 
system}

The advantage of using higher spectral resolution is 
the capability to resolve multiple components along the line of sight. 
In this instance, we identify two kinematic  
components at $z_{1a}$=1.47590 and  $z_{1b}$=1.47688 in the ESI spectrum. 
The presence of multiple components immediately 
raises the question of structure within the host galaxy.  
We might be probing a galactic disk accompanied  
by an additional kinematic component introduced by absorbing clouds 
in the halo (Disk-Halo scenario), or instead the two components 
could be due to absorption arising in distinct regions of 
a galactic disk (Disk scenario).  In order to test these scenarios
we also used information about element abundances and dust depletion. 
In 
particular, the ratios [Zn/Cr] and [Zn/Mn] provide a reasonable measure of 
the gas-to-dust ratio and the temperature of the absorbing medium.

The main line of argument lending support to 
a Disk-Halo scenario 
is the absence of a $z_{1a}$ kinematic component in 
the resolved set of Mn$^{+}$ lines. 
This is better illustrated in Figure 5, which shows 
the kinematic components for different metallic 
lines in velocity space using
$z_{1b}=1.47688$, which is assumed to be the systemic 
redshift, as the velocity zero point. 
Mg$^0$, Mn$^{+}$ and Mg$^{+}$, are present in the $z_{1b}$ system,
while the $z_{1a}$ component lacks Mn$^{+}$ but contains
Mg$^{+}$ and some of the structured but likely saturated
Fe$^{+}$ lines. 
Since the ionization potential of Mn$^{+}$
is almost identical to that of
Mg$^{+}$, a possible reason for the absence of Mn$^{+}$ at $z_{1a}$ 
is an underabundance of Mn in a metal-poor environment like the halo. 
This is based on the conjecture that Type Ia supernovae
are an important source of Mn (Samland 1998, Nakamura et al. 1999),
or that there is a
metallicity dependence yield of Mn in massive stars (Timmes, Woosley,
\& Weaver 1995), as
discussed in the case of damped Ly-$\alpha$ systems by Pettini et al. (2000). 
An early appearance of type Ia SNe during the halo phase
in Milky Way formation models has been questioned
(Chiappini, Mateucci, \& Romano 2001).
If the latter authors are correct, our
$z_{1a}$ component might arise in a halo cloud, separated 
from the velocity zero point by $\approx -119$ km~s$^{-1}$, 
with signatures of Mg$^{+}$ and Fe$^{+}$ but underabundant in Mn$^{+}$.

The possibility of a Disk-Halo line of sight makes the abundance ratios 
for each kinematic component that much more crucial in probing the
ISM at the host galaxy. Zinc
is a fairly reliable metallicity indicator since it resides more than
any other metal in the gas phase. On the other hand, metals like 
Cr and Mn are more easily depleted onto grains and a large fraction of their 
abundance is contained in metallic dust grains in the galaxy. If the 
$z_{1b}$ component is an absorption feature of the halo then a higher 
gas-to-dust ratio than in the disk is expected. Unfortunately, 
the [Zn/Mn] ratio in the $z_{1b}$ component is only an upper limit
given that Zn$^{+}$ lines from both velocity
components are blended, as are the Cr$^{+}$ lines. 
We note however that the Mn and Cr abundances are comparable. 
In fact the observed [Zn/Cr] $\approx 0.61$ is less than  
1.61 $\geq$ [Zn/Cr] $\geq$ 1.04 observed towards $\zeta$ 
Oph (Savage \& Sembach 1996) and 1.85 $\geq$ [Zn/Cr] $\geq$ 0.75
measured in a sample of 20 sightlines observed with HST (Roth \& 
Blades 1995). Although the [Zn/Cr] ratio 
is less than in any measured  Galactic sightline it does not 
represent a clear discriminant 
between a halo and a disk or a damped Lyman absorber since the spread in 
[Zn/Cr] is large for either sample (Roth \& Blades 1995).

Our results might be an indication of low dust content but not at the extreme
reported by  Salamanca et al. (2001).
It seems that their Zn II measurements
are systematically lower than those presented here or 
any other published results for GRB 010222 (see
Jha et al. 2001, Masetti et al. 2001). In
addition, Zn II(2026.14 \AA)
listed at  $\lambda_{obs} = 5011.2$ \AA~in Table 2 of Salamanca et al. (2001)
appears to be a misidentification. 
We find that line
is more consistent with Fe II(2600.17 \AA) at the $z_{3}$ system 
while Zn II should be around
$\lambda_{obs} = 5018.98$ \AA~at the higher redshift. For completeness 
we mention that we have not detected any of the unidentified 
lines listed in 
Table 3 of Salamanca et al. (2001) in the region covered by our spectra. 
Finally, the reduced [Zn/Cr] ratio in the spectrum
might be hinting that we have detected a 
warm medium in the $z_{1b}$ component   
where depletion for Cr and Mn converge (Savage \& Sembach 1996) or else
that there is little dust to deplete onto in this environment. A 
warmer medium would result in different dust depletion corrections 
and lower the inferred values of $N_{\rm HI}$ assuming standard solar 
abundances. These results are in agreement with the depletion patterns
discussed by Savaglio et al. 2002.

We now consider the alternative of a Disk scenario. In general, any neutral 
Mg present in a galactic disk must be 
shielded from external photoionizing flux. A feasible scenario
is that we are probing a line of sight passing through two distinct 
regions within the disk containing neutral gas.
These regions could be underdepleted molecular clouds 
cocooning Mg in its neutral form. 
A possible consequence of uneven photoionization in the disk is the marginal
detection of a blue ``wing'' on the $z_{1a}$ component of the 
Mg II (2796.35 \AA) absorption line (Figure 5). This feature would argue for a 
stronger photoionizing source near the closer side of the disk and  
against a galactic halo.  
This option would eliminate the necessity for a halo component, but
it does not explain the absence of Mn$^{+}$ in the $z_{1a}$ system,
assuming similar enrichment processes for different parts of the disk.
Instead, there may be a strong 
metallicity gradient along the disk, as has been observed
along the disk of the Milky Way (van Steenberg \&  Shull 1988). 
Assuming that a Disk scenario is correct and that the [Zn/Mn] ratio 
peaks towards the inner part of the host galaxy; the velocity systems 
might be tracing a fraction of galactic rotation.
Independent of the interpretation, the presence of a 
galactic disk appears necessary to explain the 
properties of the strongest absorption
system. This furthers the connection of long-duration GRBs with a 
disk progenitor.  

\subsection{Looking for signatures of $H_{2}$ absorption and Fluorescence}

The impact of the associated UV flash on $H_{2}$ molecules in the 
vicinity
of a GRB has been determined by Draine (2000). The direct observable 
consequence of strong UV radiation from a GRB on the surrounding $H_{2}$
is the production of vibrationally excited
levels which could create strong line absorption in the
region 912 \AA\ $\leq$ $\lambda_{\rm rest}$ 
$\leq$  1650 \AA, and reradiated fluorescent 
emission in a similar range. For a redshift of 
z = 1.477 this corresponds to the region $\lambda_{\rm obs}$ $\leq$ 
4087~\AA.
Unfortunately the region where the absorption effect is 
strongest at $\lambda_{\rm obs} \approx 3096$ \AA\ falls outside our 
spectral coverage. Thus, we are only able to probe
a fraction of the absorption region near the blue end of
the spectrum where the quantum efficiency is low and the noise is high. 
The strength of the absorption depends on the amount of 
$H_{2}$ available for vibrational excitation. At the LRIS resolution 
we find no definite evidence of $H_{2}$ absorption for 
$\lambda_{\rm obs}$ $\leq$ 4087 \AA~to make a proper determination 
of $N_{H_{2}}$. However, our spectra probably rule out 
excited states of $H_{2}$ at $N_{H_{2}}$ $\geq$ $10^{20}$ cm$^{-2}$
based on the transmission plots of Draine (2000). 

\subsection{Absorption-line variability}

If the line absorption occurs sufficiently close to the burst site,
the afterglow emission
could manifest itself through photoionization of the GRB environment,
thus, variability of the absorption features.
To detect this time-dependent effect various
snapshots of the absorption spectrum are required. In this 
particular instance
we take advantage of the low-resolution spectra obtained over two consecutive 
nights to search for the predicted effect. The measured 
equivalent widths for each night as well as the statistical uncertainties 
are listed in Table 1. Comparison between the LRIS equivalent widths 
on consecutive nights shows no strong evidence for time dependence
of the absorption lines 
within statistical uncertainties. 
There is also an overall agreement
with the ESI measurements. We deduce that discrepancies in any two
measurements
are mainly due to uncertainties in the continuum level in the vicinity
of each line. There does seem to be a systematic trend of  
larger equivalent
widths on the second night below 5403.80 \AA. 
However, we also note that it is clear from the spectrum that 
the equivalent widths of lines on the blue side are relatively ``weak'' with
respect to well detected lines on the red side where such trend 
is not detected. Hence,
at this point we cannot attribute 
the systematic trend entirely to a physical effect since the relative line 
flux measurements might have been affected by the continuum fitting 
during the second night as discussed in \S 3.1 and \S 3.2. 

Although the equivalent width depends on the overall 
characteristics of the continuum, its main uncertainty is defined locally
by the error in determining 
the proper level of the continuum on either side of the line of interest. As 
discussed in \S 3.1 the errors quoted for equivalent widths 
correspond to a Poisson model of the noise in the region of interest.
Such an assumption might be oversimplified and not account for true variations
in the continuum, but we believe that we safeguard against strong systematic
effects by using the same, consistent method in all measurements. 
Hence based on our analysis we do not find statistically significant 
changes in our measurements. Taking this comparison further there 
are published equivalent width measurements for this burst elsewhere 
(Jha et al. 2001, Masetti et al 2001, Salamanca et al. 2001). Although 
the duration of observation, 
spectral resolution and brightness of the burst varies in each case, 
the reported measurement 
techniques are similar. Direct comparison finds that our results
are consistent with previous measurements. Nevertheless    
some differences are present in a few cases such as C IV, Cr II and Zn II
that were discussed in \S 4.3. 
However, we do not find systematic and statistically significant 
changes as a function of time in the published data  
that should be present if either photoionization or dust destruction
in the circumburst medium are actively dominant  
for similar species. 
We refrain from reporting a measurement for 
Al III(1862.79 \AA) 
during the second LRIS spectrum
in Table 1 because of the reduced signal-to-noise in
that region of the spectrum. For completeness we note that the
line has not disappeared and it is clearly detected in the ESI spectrum
obtained at a comparable time (Table 2). 

\section{Photoionization Code}

In order to interpret our observations under contraints imposed by
the lack of absorption-line variability, we simulated the GRB environment using
a standard photoionization model. The photoionization evolution has been
described by
Perna \& Loeb (1998) and B\"ottcher et al. (1999). 
The photoionization rate can be written in the following form:

\begin{equation} 
-{1\over n_j(r,t')}\frac{dn_j(r,t')}{dt'}=\int_{\nu_{0,j}}^\infty d\nu'\;
\frac{F_{\nu'}(r,t')}{h\nu'}\;\sigma_{{\rm bf},j}(\nu')\;,
\end{equation}

\noindent
where $n_{j}$ is the density for species $j$, $F_{\nu'}(r,t')$ is
the flux at radius $r$, $t'$ is the elapsed time in the GRB frame and 
$\sigma_{{\rm bf},j}(\nu')$ is
the photoionization cross section at frequency 
$\nu'$ using the analytical fits provided by Verner et al. (1996). We 
neglect recombination processes since the densities we will be considering
are typically $n \leq$ $10^{6}$ cm$^{-3}$, where the corresponding  
recombination timescales are much longer than the duration of the GRB and 
our observations (Perna \& Loeb 1998).
In addition the full effect of dust destruction and the return of depleted 
metals into the gas phase has been ignored in the code.

Next we constructed the input flux $F_{\nu'}(r,t')$
for this particular 
afterglow from the
observed time decay (Stanek et al. 2001) and the measured optical spectral 
index. The function has two forms to accommodate the observed
break in the light-curve at~$t_{\rm break}$=0.72 day. Thus

for t $\leq$ 0.72 day 

\begin{equation} 
F_{\nu'}(r_0,t')= 5.26 \times 10^{-28}  
\left({\nu' \over 6.94 \times 10^{14}(1+z) {\rm Hz}}\right)^{-0.89}
\left({d^{2}\over (1 + z)r_{0}^{2}}\right) 
\left({t'(1+z)\over 0.352\, {\rm day}}\right)^{-0.80}
{\rm {ergs~cm^{-2}~sec^{-1}~Hz^{-1}}}
\end{equation}

otherwise 

\begin{equation} 
F_{\nu'}(r_0,t')= 2.97 \times 10^{-28}  
\left({\nu' \over 6.94 \times 10^{14}(1+z) {\rm Hz}}\right)^{-0.89}
\left({d^{2}\over (1 + z)r_{0}^{2}}\right) 
\left({t'(1+z)\over 0.72\, {\rm day}}\right)^{-1.30}
{\rm {ergs~cm^{-2}~sec^{-1}~Hz^{-1}}}
\end{equation}

\noindent
where $d$ is the luminosity distance to the burst at 
$z=1.47688$ (assuming $H_0 = 65$ km~s$^{-1}$~Mpc$^{-1}$, $\Omega_m = 0.3$, 
$\Lambda = 0.7$), and $r_{o}$ = $10^{17}$ cm is the inner radius of
the photoionized region set as a boundary condition.
We have used the observed flux and ignored any intrinsic 
dust corrections. In a previous section we showed that intrinsic
reddening following a Calzetti et al. (2000) extinction law for starbursts 
would dim and stepeen an intrinsically flat spectrum. The effect of 
using an input flux uncorrected for intrinsic extinction is to slow down 
the time-dependence in our models; therefore, the integration times
presented here should be considered upper limits.  
The input flux has been integrated in the UV range up to 
a cutoff energy of 0.2 keV. The reasoning behind this choice is partly 
arbitrary given the 
failure to locate a precise spectral break at the ``cooling frequency''  
$\nu_{c}$ between the optical and X-ray ranges. 
The selected cutoff energy of 
0.2 keV represents roughly the region where the 
photoionization cross section for the species of interest has dropped 
sharply. Above 0.5 keV, X-rays are mostly absorbed by heavy elements.

We start the photoionization of
a pristine neutral medium at initial time $t'_{o}$ that  
follows the detection of prompt emission in the BeppoSAX energy 
band. The 
transition between the prompt GRB emission and the onset of the afterglow 
is not a very well defined point observationally. Variability arguments  
set the scale for internal shocks to take place in a region of size 
$R \approx 3 \times 10^{14}$ cm $(\delta t/1\, {\rm s})(\Gamma/100)^{2}$.
External shocks responsible for the afterglow become 
significant at $\approx 10^{16}$ cm from the source (Piran 1999).
For numerical reasons we have selected an 
initial time of $t'_{o}$ = 10 seconds for the 
onset of photoionization.

The photoionization process can be pictured as ocurring inside 
a sphere of outer radius $r_{\rm max}$, subdivided into thin spherical 
shells of width $\Delta r$ so that each shell $i$ is optically 
thin. In each shell the initial incident flux spectrum described above will be 
absorbed and will also follow a $r^{-2}$ law as r $\rightarrow$ $r_{\rm max}$.
Hence the incident spectrum at any shell $i$ will be given by: 

\begin{equation}
F_{\nu'}(r_{i+1},t') = F_{\nu'}(r_{i},t') e^{-\tau_{\nu,i}}
\left({r_{i} \over r_{i+1}}\right)^{2}
\end{equation}

\noindent
We ignore bound-bound processes therefore 
$\tau_{\nu,i}$ stands for the photoionization optical depth 
which is estimated within each shell $i$ as:

\begin{equation}
\tau_{\nu,i} = \Delta r \sum_{a} n_j(r,t') \times \left[\sigma_{{\rm bf},a}(\nu')\right] 
\end{equation}

\noindent
where $\tau_{\nu,i} << 1 $. For our models 
we included H, He, and Mg and neglected other elements. 
The initial number density $n_{j}$ for each species 
is largely unknown but we decided to use 
standard solar-system abundances. 
For simplicity we also assumed that initially Mg is 
equally divided into Mg$^0$ and Mg$^{+1}$.

\section{Numerical Results \vs~Physical Conditions}

Since the depletion correction/abundance and therefore the total $N_{\rm H}$
near the burst site is not well determined,
we examined the time dependence of equivalent widths for 
two different hydrogen column 
densities. The first, log$(N_{\rm HI}) = 20.44$ cm$^{-2}$
represents the inferred 
HI column ignoring any dust depletion correction and takes into account
the possibility of complete circumburst dust destruction. The 
second, log$(N_{\rm HI}) = 22.00$ cm$^{-2}$
is appropriate for standard Milky Way depletion 
onto dust or perhaps less than solar metal abundances. The latter is also
consistent with measurements 
of the red wing of a possible Ly$\alpha$ feature (Salamanca et al. 2001), and 
the BeppoSAX afterglow results (in't Zand et al. 2001). 
In each case 
we chose four different combinations of number 
densities $n_{\rm HI}$ and $r_{\rm max}$ to 
reproduce the simulated hydrogen column density
$N_{\rm HI}=n_{\rm HI}\times r_{\rm max}$. The 
parameters for each model are listed in Table 5. 
The denser, more compact models represent the conditions in  
dense molecular clouds  
which have been postulated as a potential ambient environment for the 
progenitors of 
GRBs. On the other hand, the low density and extended environments are 
compatible with low density molecular clouds, young stellar clusters, and
possibly galactic superbubbles (Scalo \& Wheeler 2001).

We considered it relevant to trace 
the time evolution 
of two particular absorption lines present in our spectra, 
Mg I (2852.96 \AA) and Mg II (2796.35 \AA).    
Mg I is an appealing line given 
the possibility of encountering Mg in its neutral form 
in regions of the galaxy shielded from photoionization. 
Mg II(2796.35 \AA) is of interest because of its detection in 
a number of GRB spectra.  An additional observational reason for choosing 
these particular lines
to test our models is the consistency of the line flux
measurements on the red side of the spectrum where the effects of differential
refraction appear of less importance. 
In order to model our observations, we integrated
to a maximum time $t_{\rm max}=32.45$ hours after the GRB,
when the second 
LRIS spectrum was obtained. The equivalent width at different 
stages was estimated using equations (1) and (2) which
return $W_{\lambda}$. Throughout our
calculations we have assumed a
constant Doppler parameter $b$ representative of the system (see 
Table 3).  This appears to be a fairly robust assumption since 
the time dependence for the lines is mainly a function of the 
photoionization equations. 

Figure 6 illustrates the time evolution of the 
equivalent width for Mg I (2852.96 \AA) and Mg II (2796.35 \AA) in a
log$(N_{\rm HI})= 20.44$ cm$^{-2}$ environment. The times when the LRIS spectra
were obtained are indicated. 
Notice the distinct differences in 
the line evolution in each case. The 
denser, more compact models are ionized rather rapidly since
most of the gas is close to the GRB and gets
impacted by a large flux. On the other hand, less dense models 
representative of thin galactic environments require 
longer for noticeable effects to take place. The $n_{\rm HI}=89$ 
cm$^{-3}$, $r_{\rm max}=1$ pc model appears to mark the line where 
our low-resolution 
spectroscopy would have been able to detect time dependence with this 
column density. Denser environments than this
appear to be disfavored according to our model.

Figure 7 shows the time dependence for a log$(N_{\rm HI})= 22.00$
cm$^{-2}$ environment. 
Similarly compact, high density environments are photoionized rather quickly 
while extended, less dense ones would evolve with a shallower decline. 
The $n_{\rm HI}=2 \times 10^4$
cm$^{-3}$, $r_{\rm max}=5.0 \times 10^{17}$ cm model is close to 
our detection limit for this denser environment.
The results in this case are more 
dependent on the choice of the inner radius
$r_{o}$ because the scale of some of the models becomes 
comparable to $r_{o}$. A different choice for the inner radius
would either accelerate or delay the photoionization 
timescale, therefore, compact models with this column density are more 
tentative. However, it appears that 
very dense, compact environments would have been more highly ionized
than is observed in our initial spectrum. 

One interpretation of X-ray and optical afterglow data from
GRB 010222 invokes a
transition from a relativistic to a non-relativistic regime, and 
consequently the deceleration of the shock, to account for  
the break in the light curve (in 't Zand et al. 2001). This scenario 
requires an environment with $n_{\rm HI} \approx 10^{6}$ 
cm$^{-3}$ and size $\approx 10^{16}$ cm. Although our simulations argue
against the presence of a dense $10^{16}$ cm neutral gas slab 
between the two observations, we cannot 
rule out the presence of 
a large density slab at the time of the burst itself. However our
modelling also presents an alternative scenario for more extended $N_{H}$ 
absorption in the host galaxy rather than a single slab. 
Recent modelling of the multiwavelength
observations of  GRB 010222, including the radio emission, finds that
a jet model expanding into 
homogeneous external medium with density $n_{\rm HI} \approx 3.2$ cm$^{-3}$
is consistent with the broadband data (Panaitescu \& Kumar 2001). 
A jet and
low-density interpretation is also required to satisfy the standard-energy
output with beaming corrections of Frail et al. (2001a). 
One final point in this discussion is 
the shape of the $\gamma$-ray burst light curve
and baryon contamination. A very dense 
nearby medium gives rise to 
high optical depth $\tau$ where baryons could prevent relativistic expansion. 
The overall effect could lead to ``dirty'' fireballs and smearing
of the $\gamma$-ray profile. Instead the burst profile appears highly 
structured (in 't Zand et al. 2001).

Consider now the actual column density to this burst 
if the host galaxy has a dust-to-gas ratio similar to that of the
Milky Way.  A color excess
of $E(B-V) = 2.00$~mag is estimated for log$(N_{\rm HI}) = 22.00$ cm$^{-2}$
and assuming a Galactic conversion
$N_{\rm HI}/E(B-V) = 5.0 \times 10^{21}$~cm$^{-2}$~mag$^{-1}$ (Savage \& 
Mathis 1979). This is significantly larger than $E(B-V) = 0.06$~mag for 
a  log$(N_{\rm HI}) = 20.44$ cm$^{-2}$ environment. 
It is possible that the UV emission from the burst is responsible for 
destruction of neighboring dust in the high-density case
(Galama \& Wijers 2001), but further extinction at larger distances 
is expected along the line of sight unless the burst is near the 
Earth-facing side of the host galaxy (Reichart \& Price 2002). 
We must keep in mind, however, the 
possibility that we are probing different scales in two separate energy 
bands. 

\section{Implications for the Nature of the GRB 010222 Progenitor 
Environment}

GRB 010222 is a good candidate to explore whether we  
can place constrains on the progenitor environment. 
The kinematic components resolved in the high-resolution
spectrum give support to a location of this burst in
a galactic disk. High-resolution spectroscopy of GRB 000926 (Castro 
et al. 2001b) favors a similar interpretation for that burst.
The slope of the spectral continuum in this instance suggests that the 
burst occurred in a dust-free environment or was surrounded by a screen  
of gray dust (Aguirre 1999). Near-IR and submillimeter observations 
by Frail et al. (2001b) also appear to rule out the possibility of strong 
dust scattering or thermal emission near the GRB site. 
However it is difficult to model {\it a priori} the properties of dust at 
this redshift since
the distribution and properties of 
galactic dust as a function of redshift are largely unknown.

\subsection{Isolated Molecular Clouds}

Popular models of gamma-ray burst progenitors include young massive 
stars (Woosley 1993, Vietri \& Stella, 1998). This makes the 
interior of molecular 
clouds a potential birthplace for GRBs.  Molecular clouds tend to 
comprise large amounts of 
dust and clumpy pockets of gas ideal for star formation. 
If GRB 010222 took place in a molecular cloud, 
the slope of the optical continuum indicates that 
either the nearby dust was destroyed rather immediately by the burst through 
a combination of sublimation by an optical/UV 
flash (Waxman \& Draine 2000) and grain charging (Fruchter, 
Krolik \& Rhoads 2001) or that the surrounding 
dust content was low prior to the explosion. The absence of measurable 
signatures of metals being returned to their gas phase as a by-product of
dust destruction on such a short timescale does not allow direct 
distinction between these two possibilities. However, 
a high fraction of dust results 
in a larger $H_{2}$ formation rate since molecular hydrogen
tends to form in reactions taking place on grains. 
The absence of strong absorption 
features due to vibrational excited levels of $H_{2}$ expected from 
strong UV emission impacting the 
surrounding $H_{2}$ that might permeate a progenitor in a dusty molecular 
cloud (Draine 2000) is conspicuous. 
Moreover, assuming a mean density
of $10^{4}$ cm$^{-3}$ and the typical size of a dense molecular cloud, 
the absorption lines detected should have weakened over the period of our 
observations. A massive and dense molecular cloud might also give rise
to reflection at 6.4 keV as seen near the Galactic center (Murakami, Koyama,
\& Maeda 2001).  Nevertheless, it is still viable that a molecular hydrogen
detection has been made blueward of our spectral limit by Lee et al. (2001) 
and Galama et al. (2002). 
As reported by these authors the flux deficit in the $U$~
band
which extends below 3500 \AA\ does not fit the SMC or LMC-like extinction 
models. A depression signature is also expected if there is absorption
by a Lyman-$\alpha$ wing (Salamanca et al. 2001). The absence of molecular 
hydrogen and reduced depletion resulting from the [Zn/Mn] and [Zn/Cr] ratios
still allows the location of the GRB to be near the edge of a molecular cloud
prior to the explosion.
Thus, at the time of the event the fraction of gas 
to dust can be larger than at the center of the cloud, 
inhibiting the formation of $H_{2}$. 
Immediately after the onset of 
the UV emission from the burst, any remaining dust could have been destroyed 
efficiently by the expanding UV radiation pulse, minimizing reddening of the 
optical continuum.

\subsection{Superbubbles}

Panaitescu \& Kumar (2001) in their multiwavelength analysis of 
well-studied afterglows found that the ambient density for some bursts 
can be as low as $1.9 \times 10^{-3}$ cm$^{-3}$; however, 
some arguments favor higher ambient densities once the complete radio data
set is included in the analysis of individual bursts (Berger et al. 2002). 
The presence of  
low densities led Scalo \& Wheeler (2001) to argue for
active star formation regions in 
superbubbles as possible environments for GRBs. Our 
photoionization models show that low-density environments with the typical
sizes of superbubbles would not produce strong time-dependence
in absorption lines.  Moreover, the interior of a superbubble  
could be partionally ionized prior to 
the GRB event. The prior ionization might be accomplished 
via overlapping ionization bubbles generated by supernovae
in clusters of massive stars. A pre-ionized environment decreases 
the amount of
neutral material seen by the expanding photoionization front and 
prevents the full effect of time-dependent absorption.
But it also enables the detection of
time dependence in high-ionization species. We detect high-ionization 
C IV absorption in our low-resolution spectrum, but it is not clear that 
equivalent width variations are seen. In general
C IV is a tracer of additional diffuse warm and hot gas that 
could originate in a 
superbubble environment. Unfortunately, our high-resolution spectrum does
not cover this absorption line to firmly establish if it comes from the
disk of the host galaxy or rather from a gaseous halo. We note however that 
the equivalent width of  C IV (1548.20 \AA) is comparable to that of the Mg II
absorption lines. An additional signature of prior ionization could be
the attenuation of the X-ray continuum by an ionized absorber like 
those seen in AGN spectra. The absence of strong absorption features except
for a marginal Fe XIV-XVIII edge suggested by in 't Zand et al. (2001)
prevents a straightforward connection from being made between X-ray and
UV absorption features.

An advantage of the superbubble scenario is 
the availability of a  dust destruction mechanism
and a favorable geometry for observation. 
The combination of winds and SN explosions would tend to blow 
the gas and dust away leaving filamentary or shell-like patterns like the one 
reported in the Galactic Orion-Eridanus bubble (Guo et al. 1995). 
In addition, winds and radiation pressure 
in the interior of a superbubble might selectively destroy small grains, 
a 
mechanism for gray dust production locally. 
However, this selective
process is balanced by rapid destruction of large grains by the burst and 
earlier destruction by X-rays emitted inside the superbubble that might 
lead to grain charging (Fruchter, Krolik, \& Rhoads 2001). In combination, 
these
two processes could counteract the expulsion of small grains  
and suppress the formation of gray dust.  Either dust destruction or 
its gray nature are consistent with the slope of the observed continuum 
and the lack of strong
dust reprocessing near the burst site (Frail et al. 2001b). 
The only practical problem with this interpretation 
is the mismatch between the optical and
X-ray spectra.  It is possible that the cooling frequency has moved
out of the optical band.  In this case we propose that the observed X-ray 
excess (in 't Zand et al. 2001) is mostly due  
to inverse Compton scattering rather than intrinsic reddening near the
burst location but if true the ambient density cannot be 
below a few $cm^{-3}$ or inverse Compton scattering would be 
suppressed (Harrison et al. 2001).  If the effects of dust
depletion are less pronounced than in the Milky Way, a 
log$(N_{\rm HI}) = 20.44$~cm$^{-2}$ environment with
$n_{\rm HI} \leq 89$~ cm$^{-3}$ and $r_{\rm max} \geq 1$~pc is suggested 
for GRB 010222. This is consistent with the best-fitted parameter
$n_{\rm HI} \approx 3.2 $~cm$^{-3}$ resulting from
jet modelling (Panaitescu \& Kumar 2001). 

\subsection{Young Stellar Clusters}

The recently discovered Arches cluster (Nagata et al. 1995; Cotera et al. 1996;
Serabyn, Shupe, \& Figer 1998; Figer et al. 1999) consists of more that 100
young stars with masses greater than 20 \Msol~ within a $\approx 0.3-0.4$~pc
radius. Radio observations of the Arches stellar cluster have confirmed
the presence of powerful ionized winds with mass-loss rates $\approx
(1-20) \times 10^{-5}$ \Msol~yr$^{-1}$ coincident with massive 
stars in the cluster (Lang, Goss, \& Rodriguez 2001).
If a low metallicity massive star
that has lost its envelope turns out to be the prototype for  
the collapsar model of GRBs (MacFadyen \& Woosley 1999), a young
stellar cluster environment
appears to be a perfect cradle for progenitors as prelude 
to GRB formation.  A compact young cluster 
environment is consistent with our observations
and models but requires low dust content or rapid destruction 
near the burst site, as well as photoionization of the gas in the 
region by hot young stars prior to the GRB.   
The complexity of wind interaction and shocks 
might provide favorable
dust viewing geometries and dust distribution similar to those
discussed in the superbubble scenario. In addition,
the UV absorption lines and their lack of time dependence 
might be attributed to the 
medium well outside the stellar cluster where the photoionizing flux 
is not strong enough to imprint substantial changes. 
One problem with this interpretation is the type of metallicities required
by the MacFadyen \& Woosley (1999) model which can be thought
of as LMC-like (A.I. MacFadyen 2002, private communication). The 
abundances in the GRB 010222 
host are larger than the abundances reported in LMC line of sights
(Welty et al. 1999). In addition, the absence of 
strong signatures of ionized absorption 
and pronounced emission lines in the X-ray spectrum (in't Zand et al. 2001) 
might argue against a pre-ionized medium. 

\subsection{Halo}

An alternative GRB progenitor is a coalescing 
compact object (Neutron Star-Neutron Star, Black Hole-Neutron Star)
(Paczy\'nski 1986). Even 
though the distribution of birth kick velocities for compact 
objects has not been fully surveyed, 
typical velocities around a few hundred km s$^{-1}$ could place a 
fraction of potential progenitors at high-latitude above the disk.
The apparent presence of a galactic disk in this case 
seems to rule out the possibility of
a halo origin for GRB 010222. Even though 
a location in the halo implies a low-density environment, the 
ionized regions of interest 
would be relatively compact clouds in the halo. 
In this case, it is difficult
to reproduce the measured neutral column densities and observed time
evolution in a halo event. Photoionization in halo events might
be more effective in erasing absorption features entirely if the
line-of-sight does not intercept the disk of the host galaxy.
Current ideas suggest that the hard-short class of GRBs are due 
to coalescing compact objects (Janka \etal~1999, Fryer \etal~1999). Rapid 
localization of the hard-short GRBs might 
allow the detection of absorption-line time dependence 
in high-ionization species following coalescence in a galactic halo.

\section{Conclusions and other Observational Reflections}

The use of high-resolution spectroscopy has resolved the host galaxy of GRB 
010222 into multiple components. We conclude that the GRB took place 
in a galactic disk that
gives rise to the strongest absorption system. The power-law index of the 
optical continuum indicates that the dust content
at the burst place is low or that
the reddening follows a gray dust model. Under these conditions the bulk 
of the X-ray excess (in 't Zand et al. 2001)
can be attributed to inverse Compton
scattering.  Low resolution spectroscopy 
obtained over the span of two days failed to reveal any significant evidence 
for time dependence of the observed absorption lines. Signatures of 
strong $H_{2}$ 
absorption and/or fluorescence are also absent from our spectra. 
Photoionization models constructed to test a range   
of initial conditions for this burst 
show that dense,  
compact media would be photoionized fairly 
quickly while less dense media with large radii 
evolve slower with time. We argue that prior to a GRB,
an environment partially ionized
either by the progenitor or by nearby sources would 
prevent the full manifestation of predicted absorption-line evolution.  
The physical attributes of GRB 010222 are consistent with a low-density 
molecular cloud, superbubble, or young stellar cluster as the possible 
environment of the GRB.  

In the future some consideration should be given to a prompt optical-UV flash 
like the one accompanying GRB 990123. The intensity of such an event
could in principle lead to dust destruction, but it might also be responsible
for the photoionization of a fraction of the neutral gas. 
The full effect of such a flash has been ignored in our models.
Preferentially, the observation of time-dependent absorption lines
would work best in
bursts occurring in the Earth-facing side of the host galaxy and 
in compact regions. 
Our analysis suggests that this technique applied with high-resolution 
spectroscopy could potentially
distinguish the metallicity of progenitor environments.
In that regard shortening 
the interval between burst localization and initial spectroscopy becomes
crucial to exploring earlier, less photoionized times.
The next generation of high-resolution 
imagers might directly resolve the environment of GRBs in a relatively 
nearby host galaxy.  A prospect raised by the observations of GRB 010222 is 
the study of the properties of galactic dust at 
high-redshift using bursts as beacons to illuminate the dust nearby.  
Further exploration of this issue is 
encouraged as spectroscopy is obtained for a larger sample of GRBs.

\acknowledgments{ }
We would like to thank R. -P. Kudritzki and F. Bresolin for obtaining the 
first 
set of spectra as well as H. Spinrad, D. Stern,
A. Dey, A. Bunker, and S. Dawson for taking the second set of LRIS spectra. 
We also acknowledge Eric Gotthelf for allowing us to use his new 
Alpha computer and Robert Uglesich for tips on optimization in FORTRAN. 
This work was supported by the National Science
Foundation under Grant No. AST-00-71108.
JSB is supported as a Fannie and John Hertz Fellow.

\clearpage

\begin{deluxetable}{cccccc}
\tablenum{1}
\tablewidth{0pt}
\tablecaption{LRIS Line Identification  }
\tablehead
{
Line($\lambda_{\rm vac}$(\AA)) & $\lambda_{\rm helio}$(\AA) & $z$ &  $f_{ij}$ &
$W_{1}$(\AA)\tablenotemark{c}
& $W_{2}$(\AA)\tablenotemark{d}
}
\startdata
Si II(1526.71 \AA)  &  3782.12 & 1.4773 & 1.10$\times 10^{-1}$ & 1.51 $\pm$ 0.13 	&  1.72 $\pm$ 0.47\\
C IV(1548.20 \AA) & 3835.36 & 1.4773 & 1.91$\times 10^{-1}$ & 2.55 $\pm$ 0.23	&  2.67 $\pm$ 0.29\\
+ C IV(1550.78 \AA)\tablenotemark{b} &  &   & 9.52$\times 10^{-2}$ &   & \\
Fe II(1608.45 \AA) & 3984.61 & 1.4773 & 6.19$\times 10^{-2}$ & 0.40 $\pm$ 0.04	&  0.52 $\pm$ 0.30\\
Al II(1670.79 \AA) & 4138.77 & 1.4771 & 1.83 & 1.48 $\pm$ 0.09	&  1.58 $\pm$ 0.12\\
Si II(1808.01 \AA) & 4478.51 & 1.4770 & 5.53$\times 10^{-3}$ & 0.60 $\pm$ 0.08	& 0.81 $\pm$ 0.15\\
Al III(1854.72 \AA) & 4592.77 & 1.4763 & 5.60$\times 10^{-1}$ & 0.79 $\pm$ 0.06 &  0.94 $\pm$ 0.13\\
Fe II(2382.77 \AA)\tablenotemark{a} & 4592.77 & 0.9275 & 3.01$\times 10^{-1}$ & 1.01 $\pm$ 0.08 &  1.21 $\pm$ 0.17\\
Al III(1862.79 \AA) & 4614.50 & 1.4772 & 2.79$\times 10^{-1}$ & 0.69 $\pm$ 0.08	&  . . .  \\
Zn II(2026.14 \AA) & 5018.95 & 1.4771 & 4.89$\times 10^{-1}$ & 0.71 $\pm$ 0.07 &  1.03 $\pm$ 0.13\\
+Mg I(2026.48 \AA) \tablenotemark{b} &  &  & 1.15$\times 10^{-1}$ &  &  \\
Cr II(2056.25 \AA) & 5093.54 & 1.4771 & 1.05$\times 10^{-1}$ & 0.3 $\pm$ 0.04	& 0.45 $\pm$ 0.08\\
Zn II(2062.66 \AA) & 5109.62 & 1.4772 & 2.56$\times 10^{-1}$ & 0.7 $\pm$ 0.04 & 0.52 $\pm$ 0.09\\
+Cr II(2062.23 \AA)\tablenotemark{b} &  &  & 7.80$\times 10^{-2}$ &  &   \\
Fe II(2382.77 \AA) & 5137.88 & 1.1563 & 3.01$\times 10^{-1}$ & 0.64 $\pm$ 0.08	&  0.81 $\pm$ 0.11\\
Mg II(2796.35 \AA) & 5390.80 & 0.9278 & 6.12$\times 10^{-1}$& 1.04 $\pm$ 0.10	&  1.06 $\pm$ 0.13\\
Mg II(2803.53 \AA) & 5403.80 & 0.9275 & 3.05$\times 10^{-1}$ & 0.71 $\pm$ 0.07	&  0.91 $\pm$ 0.08\\
Mg I(2852.96 \AA)  & 5493.01 & 0.9254 & 1.83 & 0.67 $\pm$ 0.13	& 0.63 $\pm$ 0.15\\
Fe II(2600.17 \AA) & 5605.97 & 1.1560 & 2.24$\times 10^{-1}$ & 1.32 $\pm$ 0.07 & 1.21 $\pm$ 0.11\\
Fe II(2344.21 \AA) & 5807.30 & 1.4773 & 1.10$\times 10^{-1}$ & 1.97 $\pm$ 0.06	& 1.86 $\pm$ 0.06\\
Fe II(2374.46 \AA) & 5881.78 & 1.4771 & 3.26$\times 10^{-2}$ & 1.67 $\pm$ 0.08	& 1.80 $\pm$ 0.22\\
Fe II(2382.77 \AA) & 5902.60 & 1.4772 & 3.01$\times 10^{-1}$ & 2.37 $\pm$ 0.09	& 2.10 $\pm$ 0.12\\
Mg II(2796.35 \AA) & 6030.31 & 1.1565 & 6.12$\times 10^{-1}$ & 2.07 $\pm$ 0.13	& 2.08 $\pm$ 0.16\\
\enddata
\end{deluxetable}
\clearpage

\begin{deluxetable}{cccccc}
\tablenum{1}
\tablewidth{0pt}
\tablecaption{LRIS Line Identifications (Continued)  }
\tablehead
{
Line($\lambda_{\rm vac}$(\AA)) & $\lambda_{\rm helio}$(\AA) & $z$ & $f_{ij}$ &
$W_{1}$(\AA)\tablenotemark{c} & $W_{2}$(\AA)\tablenotemark{d}
}
\startdata
Mg II(2803.53 \AA) & 6044.97 & 1.1562 & 3.05$\times 10^{-1}$ & 1.77 $\pm$ 0.14	& 1.79 $\pm$ 0.15\\
Mn II(2576.88 \AA) & 6383.63 & 1.4773 & 3.51$\times 10^{-1}$ & 0.62 $\pm$ 0.07	& 0.50 $\pm$ 0.10\\
Fe II(2586.65 \AA) & 6407.39 & 1.4771 & 6.84$\times 10^{-2}$ & 1.53 $\pm$ 0.08	& 1.5 $\pm$ 0.15\\
Fe II(2600.17 \AA) & 6441.14 & 1.4772 & 2.24$\times 10^{-1}$ & 2.48 $\pm$ 0.08	& 2.2 $\pm$ 0.13\\
Mg II(2796.35 \AA) & 6926.28 & 1.4769 & 6.12$\times 10^{-1}$ & 3.06 $\pm$ 0.04	& 3.00 $\pm$ 0.11\\
Mg II(2803.53 \AA) & 6944.62 & 1.4771 & 3.05$\times 10^{-1}$ & 2.88 $\pm$ 0.04	& 2.90 $\pm$ 0.09\\
Mg I(2852.96 \AA) & 7067.07 & 1.4771 & 1.83 & 1.22 $\pm$ 0.05 & 1.31 $\pm$ 0.09\\
\enddata
\tablenotetext{a}{Alternative identification to previous entry.}
\tablenotetext{b}{Doublet or blend between lines.}
\tablenotetext{c}{Rest equivalent width on 2001 Feb 22.66.}
\tablenotetext{d}{Rest equivalent width on 2001 Feb 23.66.}
\end{deluxetable}
\clearpage

\begin{deluxetable}{ccccc}
\tablenum{2}
\tablewidth{0pt}
\tablecaption{ESI Line Identifications}
\tablehead
{
Line($\lambda_{\rm vac}$(\AA)) & $\lambda_{\rm helio}$(\AA) & $z$ & $f_{ij}$ 
& $W_{o}$(\AA) 
}
\startdata
Al II(1670.79 \AA)  & 4138.41 & 1.47692 & 1.83 & 1.87 $\pm$ 0.43\\
Si II(1808.01 \AA) & 4478.18 & 1.47686 & 2.18$\times 10^{-3}$ & 0.76 $\pm$ 0.1\\
Al III(1854.72 \AA) & 4592.39 & 1.47606 & 5.60$\times 10^{-1}$ & 0.41 $\pm$ 0.20\\
Fe II(2382.77 \AA)\tablenotemark{a}  & 4592.39 & 0.92733 & 3.01$\times 10^{-1}$ & 0.53 $\pm$ 0.26\\
Al III(1854.72 \AA) & 4594.04 & 1.47695 & 5.60$\times 10^{-1}$ & 0.80 $\pm$ 0.18\\
Fe II(2382.77 \AA)\tablenotemark{a}  & 4594.04 & 0.92803 & 3.01$\times 10^{-1}$ & 1.03 $\pm$ 0.23\\
Al III(1862.79 \AA) & 4614.18 & 1.47703 & 2.79$\times 10^{-1}$ & 0.49 $\pm$ 0.07\\
Fe II(2600.17 \AA)  & 5011.47 & 0.92736 & 2.24$\times 10^{-1}$ & 0.41 $\pm$ 0.11\\
Zn II(2026.14 \AA) &  5018.98 & 1.47711 & 4.89$\times 10^{-1}$ & 0.91 $\pm$ 0.12\\
+Mg I(2026.48 \AA)\tablenotemark{b} &  &    & 1.15$\times 10^{-1}$ & \\
Cr II(2056.25 \AA) & 5093.10 & 1.47689 & 1.05$\times 10^{-1}$ & 0.44 $\pm$ 0.08\\
Zn II(2062.66 \AA) & 5108.62 & 1.47671 & 2.56$\times 10^{-1}$ & 0.69 $\pm$ 0.16\\
+Cr II(2062.23 \AA)\tablenotemark{b} &  &  & 7.80$\times 10^{-2}$ & \\
Cr II(2066.16 \AA) & 5117.54 & 1.47684 & 5.15$\times 10^{-2}$ & 0.27 $\pm$ 0.12\\
Fe II(2382.77 \AA) & 5138.48 & 1.15652 & 3.01$\times 10^{-1}$ & 1.12 $\pm$ 0.11\\
Mg II(2796.35 \AA)  & 5389.83 & 0.92745 & 6.12$\times 10^{-1}$&  0.96 $\pm$ 0.14\\
Mg II(2803.53 \AA) & 5404.04 & 0.92758 & 3.05$\times 10^{-1}$& 1.08 $\pm$ 0.16\\
Mg I(2852.96 \AA)   & 5498.99 & 0.92747 & 1.83 & 0.51 $\pm$ 0.21\\
Fe II(2260.78 \AA) & 5597.47 & 1.47590  & 2.44$\times 10^{-3}$ &  0.10 $\pm$ 0.08\\
Fe II(2260.78 \AA) & 5599.68 & 1.47688  & 2.44$\times 10^{-3}$ &  0.42 $\pm$ 0.16\\
Fe II(2600.17 \AA) & 5606.38 & 1.15616 & 2.24$\times 10^{-1}$ & 1.15 $\pm$ 0.16\\
\enddata
\end{deluxetable}

\clearpage

\begin{deluxetable}{ccccc}
\tablenum{2}
\tablewidth{0pt}
\tablecaption{ESI Line Identification (Continued)  }
\tablehead
{
Line($\lambda_{\rm vac}$(\AA)) & $\lambda_{\rm helio}$(\AA) & $z$ & $f_{ij}$ & 
$W_{o}$(\AA)
}
\startdata
Fe II(2344.21 \AA) & 5804.03 & 1.47590 & 1.10$\times 10^{-1}$& 0.37 $\pm$ 0.11\\
Fe II(2344.21 \AA) & 5806.33 & 1.47688 & 1.10$\times 10^{-1}$& 1.51 $\pm$ 0.23\\
Fe II(2374.46 \AA) & 5878.93  & 1.47590 & 3.26$\times 10^{-2}$ & 0.30 $\pm$ 0.06\\
Fe II(2374.46 \AA) & 5881.25  & 1.47688 & 3.26$\times 10^{-2}$ & 1.24 $\pm$ 0.10\\
Fe II(2382.77 \AA) & 5899.50 & 1.47590 & 3.01$\times 10^{-1}$ & 0.56 $\pm$ 0.09\\
Fe II(2382.77 \AA) & 5901.84 & 1.47688 & 3.01$\times 10^{-1}$ & 2.21 $\pm$ 0.16\\
Mg II(2796.35 \AA) & 6028.98 & 1.15602 & 6.12$\times 10^{-1}$& 2.49 $\pm$ 0.08\\
Mg II(2803.53 \AA) & 6044.44  & 1.15601 & 3.05$\times 10^{-1}$ & 2.14 $\pm$ 0.60\\
Mg I(2852.96 \AA)  & 6152.90 & 1.15667 & 1.83 & 0.54 $\pm$ 0.31\\
Mn II(2576.88 \AA)& 6382.58 & 1.47686 & 3.51$\times 10^{-1}$ & 0.58 $\pm$ 0.12\\
Fe II(2586.65 \AA) & 6404.29  & 1.47590 & 6.84$\times 10^{-2}$ & 0.64 $\pm$ 0.13\\
Fe II(2586.65 \AA) & 6406.82  & 1.47688 & 6.84$\times 10^{-2}$ & 1.72 $\pm$ 0.20\\
Mn II(2594.50 \AA) & 6426.18  & 1.47685 & 2.71$\times 10^{-1}$& 0.68 $\pm$ 0.12\\
Fe II(2600.17 \AA) & 6437.76  & 1.47590 & 2.24$\times 10^{-1}$ & 0.49 $\pm$ 0.05\\
Fe II(2600.17 \AA) & 6440.31  & 1.47688 & 2.24$\times 10^{-1}$ & 2.12 $\pm$ 0.19\\
Mn II(2606.46 \AA) & 6455.96  & 1.47691 & 1.93$\times 10^{-1}$ & 0.56 $\pm$ 0.12\\
Mg II(2796.35 \AA) & 6923.66  & 1.47596 & 6.12$\times 10^{-1}$& 1.63 $\pm$ 0.24\\
Mg II(2796.35 \AA) & 6926.76  & 1.47707 & 6.12$\times 10^{-1}$& 1.51 $\pm$ 0.19\\
Mg II(2803.53 \AA) & 6940.97  & 1.47580 & 3.05$\times 10^{-1}$ & 1.39 $\pm$ 0.15\\
Mg II(2803.53 \AA) & 6944.06  & 1.47690 & 3.05$\times 10^{-1}$ & 1.61 $\pm$ 0.11\\
Mg I(2852.96 \AA)  & 7063.74  & 1.47593 & 1.83 & 0.42 $\pm$ 0.24\\
Mg I(2852.96 \AA)  & 7066.33  & 1.47684 & 1.83 & 1.21 $\pm$ 0.22\\
\enddata
\tablenotetext{a}{Alternative identification to previous entry.}
\tablenotetext{b}{Doublet or blend between lines.}
\end{deluxetable}

\clearpage

\begin{deluxetable}{cccccccc}
\tablenum{3}
\tablewidth{0pt}
\tablecaption{Derived Column Densities for the $z_{1b}$ System}
\tablehead
{
Ion($j$)  &  $\lambda_{\rm vac}$(\AA) & $f_{ij}$ & $b$ (km~s$^{-1}$) &
log$(N_{j})$ & log($N_{\rm HI}$)\tablenotemark{a} & 
log$(N_{\rm HI})_{\rm corr}$\tablenotemark{b} & [Zn/$J$]\tablenotemark{c}}
\startdata
Mn$^{+}$ & 2576.88 & 0.3508 & 22.8 & 14.00 $\pm$ 0.44 & 20.47 $\pm$ 0.44 & 21.92 $\pm$ 0.44 & 0.50\\
      & 2594.50 & 0.2710 & 35.3 & 13.95 $\pm$ 0.17 & 20.42 $\pm$ 0.17 & 21.87 $\pm$ 0.18\\
      & 2606.46 & 0.1927 & 29.6 & 14.00 $\pm$ 0.20 & 20.47 $\pm$ 0.20 & 21.92 $\pm$ 0.21\\ 	        	
Mg$^{+}$ & 2803.53 &  3.05$\times 10^{-1}$ & . . . &  $\geq$ 13.88\tablenotemark{d} & $\geq$ 18.30 & $\geq$ 19.85 & $\leq$ 2.65\\
Fe$^{+}$ & 2382.77 & 3.01$\times 10^{-1}$ & . . . & $\geq$ 14.16\tablenotemark{d} & $\geq$ 18.65   & $\geq$ 20.92 & $\leq$ 2.3\\
\enddata
\tablenotetext{a}{Assuming solar abundances and no depletion.}
\tablenotetext{b}{Assuming depletion corrections from Savage \& Sembach (1996).}
\tablenotetext{c}{Average log metal abundance ratio relative to Solar value.}
\tablenotetext{d}{Using Equation (3).}
\end{deluxetable}

\clearpage

\begin{deluxetable}{ccccc}
\tablenum{4}
\tablewidth{0pt}
\tablecaption{Column Densities for Blended Lines}
\tablehead
{
Ion($j$) & log$(N_{\rm j})$ & log$(N_{\rm HI})$\tablenotemark{a}& 
log$(N_{\rm HI})_{\rm corr}$\tablenotemark{b} & [Zn/$J$]\tablenotemark{c}}
\startdata
Zn$^{+}$  & 13.60 $\pm$ 0.07 & 21.04  $\pm$ 0.07 &  21.71 $\pm$ 0.07 & . . .\\ 	
Cr$^{+}$  & 14.02 $\pm$ 0.1 & 20.37 $\pm$ 0.1  &  22.65 $\pm$ 0.1 & 0.61\\
\enddata
\tablenotetext{a}{Assuming solar abundances and no depletion.}
\tablenotetext{b}{Assuming depletion corrections from Savage \& Sembach (1996).}
\tablenotetext{c}{Log metal abundance ratio relative to Solar value.}
\end{deluxetable}

\clearpage

\begin{deluxetable}{cccc}
\tablenum{5}
\tablewidth{0pt}
\tablecaption{Photoionization Model Parameters}
\tablehead
{
$N_{\rm HI}$(cm$^{-2})$ & $n_{\rm HI}$(cm$^{-3})$  & 
$r_{\rm max}$(pc) & Model}
\startdata
$10^{20.44}$ & 0.089 & 1000.0 & I\\
$10^{20.44}$ & 0.89 & 100.0 & II\\
$10^{20.44}$ & 89 & 1.0 & III\\
$10^{20.44}$ & $2.75 \times 10^3$ & 6.48 $\times$ $10^{-2}$ & IV\\
$10^{22.00}$ & 3.2  & 1000.0 & V\\
$10^{22.00}$ & $2.0 \times 10^4$  & 0.19 & VI\\
$10^{22.00}$ & $10^{5}$ & 6.48 $\times$ $10^{-2}$ & VII\\
$10^{22.00}$ & $10^{6}$ & 3.56 $\times$ $10^{-2}$ & VIII\\
\enddata
\end{deluxetable}
\clearpage

\begin{figure}[tbp] \label{lris} \figurenum{1}
\begin{center}
\epsscale{.56}
\plotone{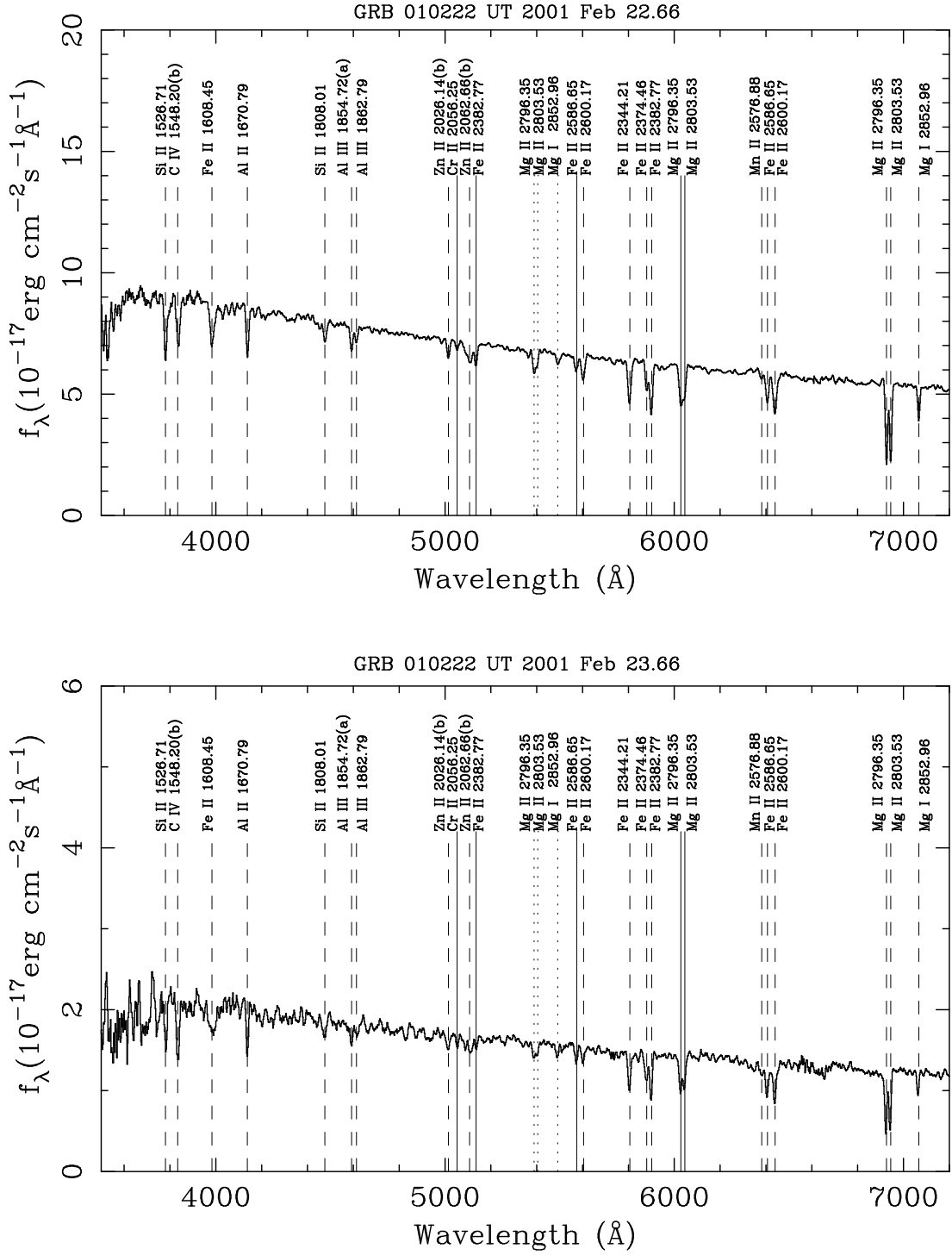}
\caption{LRIS spectra of GRB 010222 on 2001 Feb. 22.66 and Feb 23.66 
corrected for Galactic foreground extinction and 
smoothed with a boxcar  corresponding to the instrumental resolution. Three
absorption systems are labelled: $z_{1}=1.47688$ ({\it dashed lines}), 
$z_{2}=1.15628$ ({\it solid lines}), and $z_{3}=0.92747$
({\it dotted lines}). Letters in parentheses 
refer to the notes in Table 1.}
\end{center}
\end{figure}
\clearpage

\begin{figure}[tbp] \label{esi} \figurenum{2}
\begin{center}
\epsscale{.56}
\plotone{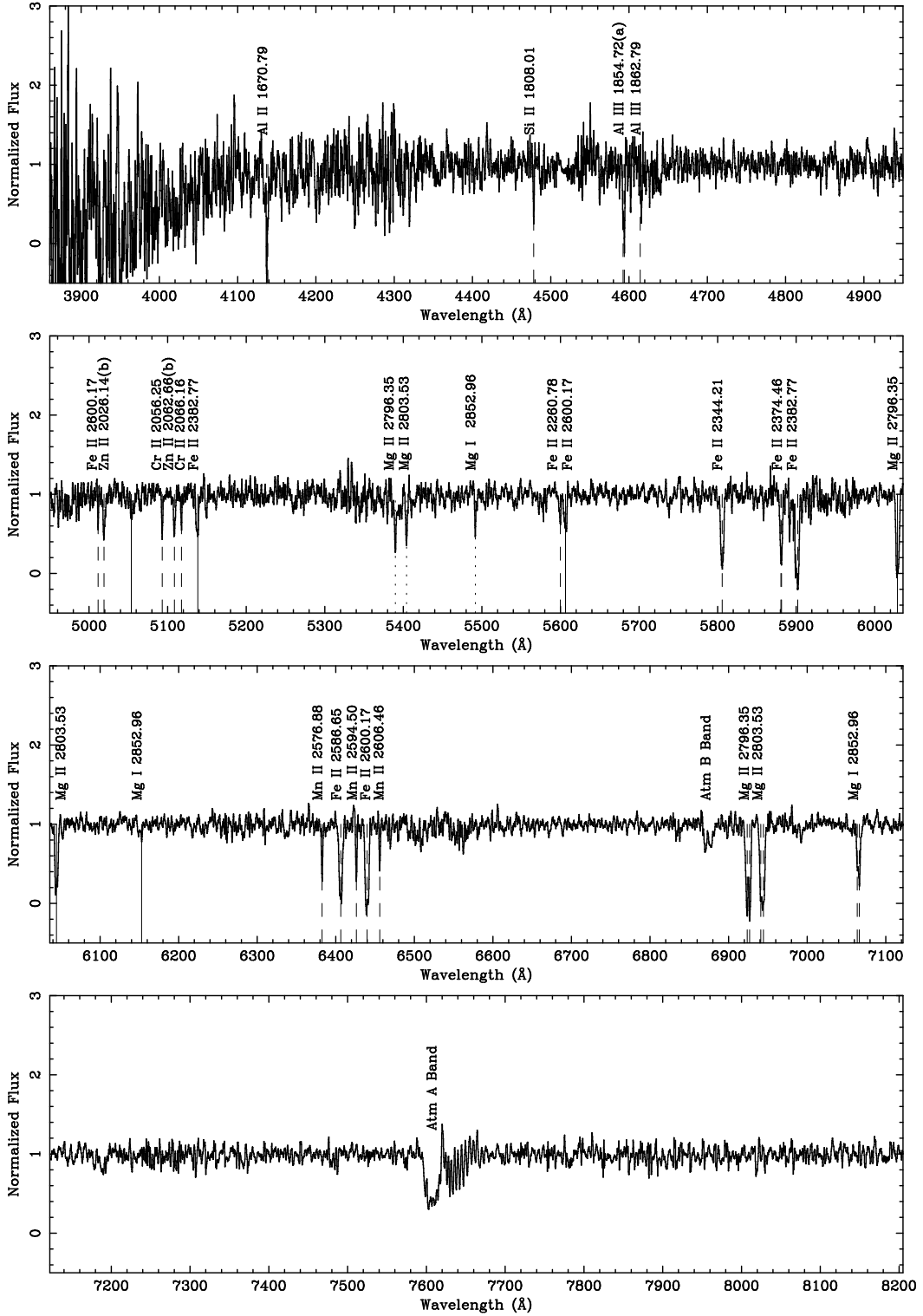}
\caption{Continuum-normalized 
ESI spectrum of GRB 010222 on 2001 Feb. 23.61. The spectrum is 
smoothed with a boxcar corresponding to the instrumental resolution. 
Three absorption systems are labelled: $z_{1}=1.47688$ ({\it dashed lines}),
$z_{2}=1.15628$ ({\it solid lines}), and $z_{3}=0.92747$
({\it dotted lines}). Atmospheric bands are also indicated. Letters 
in parentheses refer to the notes in Table 2.}
\end{center}
\end{figure}
\clearpage

\begin{figure}[tbp] \label{optcont} \figurenum{3}
\begin{center}
\epsscale{.60}
\plotone{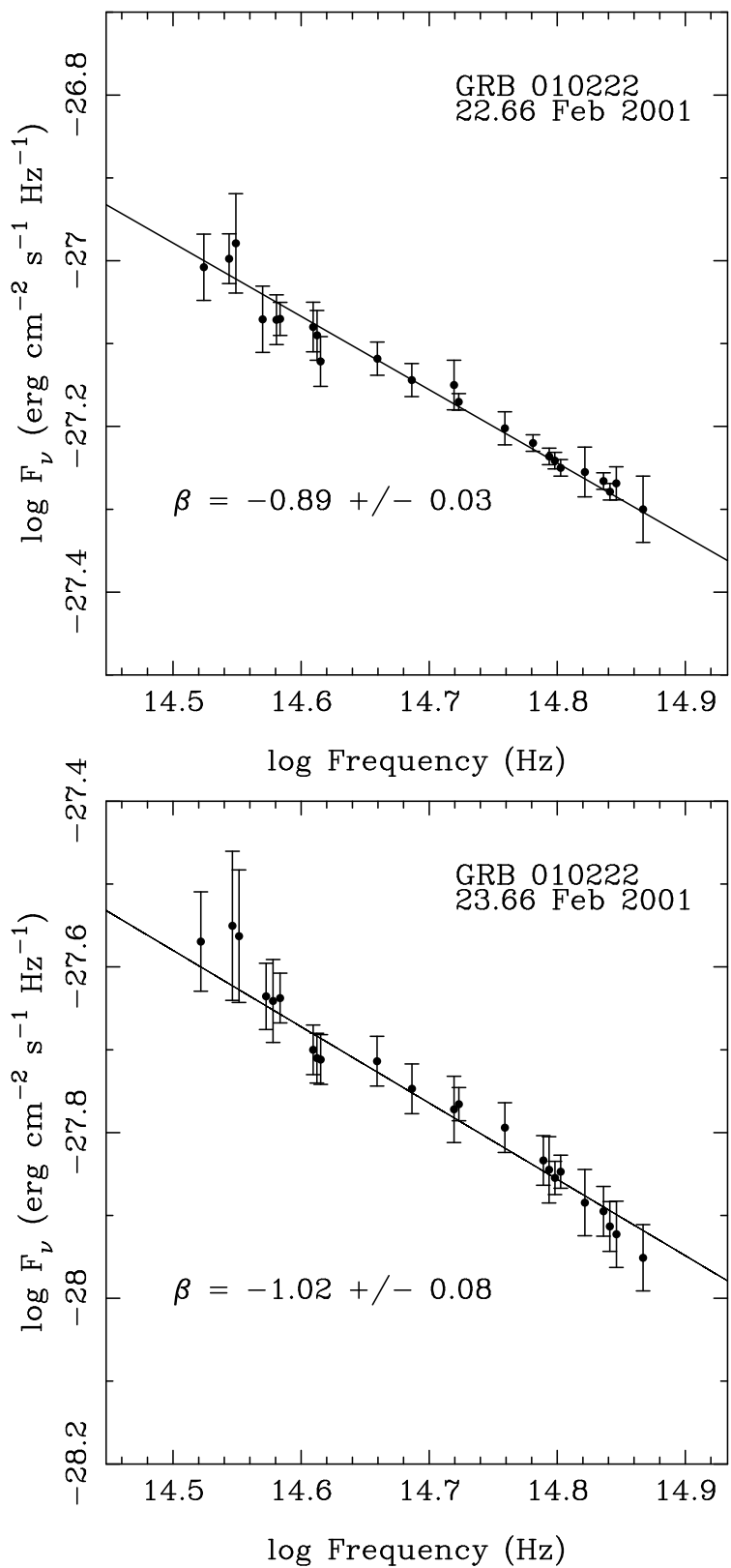}
\caption{Optical continuum points with
1$\sigma$ standard deviation uncertainties,
along with the best fitted line that minimizes $\chi^{2}$, for 2001 Feb 22.66
and 2001 Feb 23.66.}
\end{center}
\end{figure}
\clearpage

\begin{figure}[tbp] \label{multilam} \figurenum{4}
\begin{center}
\plotone{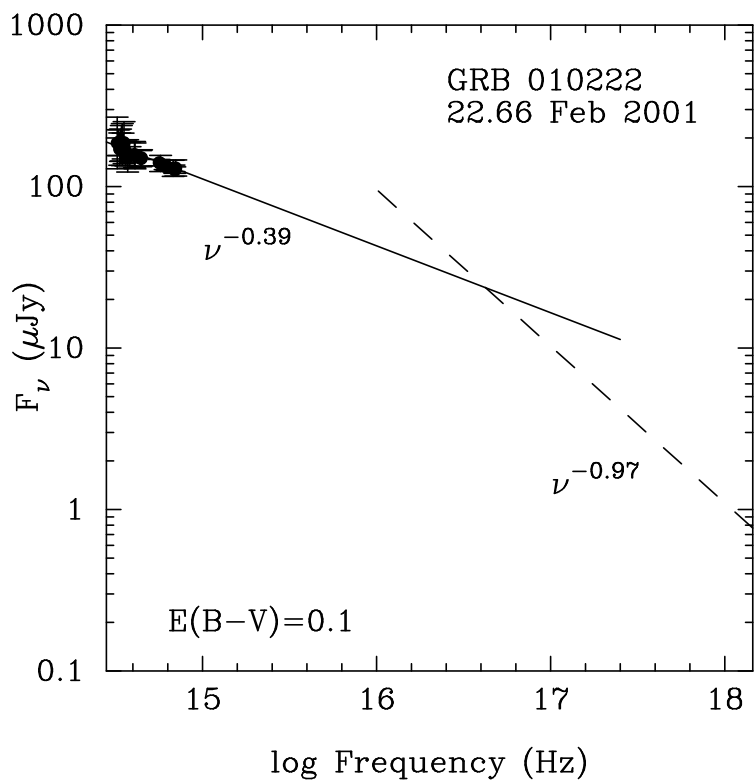}
\caption{
Multiwavelenth X-ray and optical continuum
corrected for dust using the extinction curve
of Calzetti et al. (2000) and color excess $E(B-V)=0.1$
for 2001 Feb 22.66. The
resulting optical spectrum has a slope $\beta \approx -0.43$. The X-ray 
data is shown schematically as a
{\it dashed line}, while the best fit through the optical
data is plotted as a {\it solid line}.}
\end{center}
\end{figure}
\clearpage

\begin{figure}[tbp] \label{metals} \figurenum{5}
\begin{center}
\epsscale{.58}
\plotone{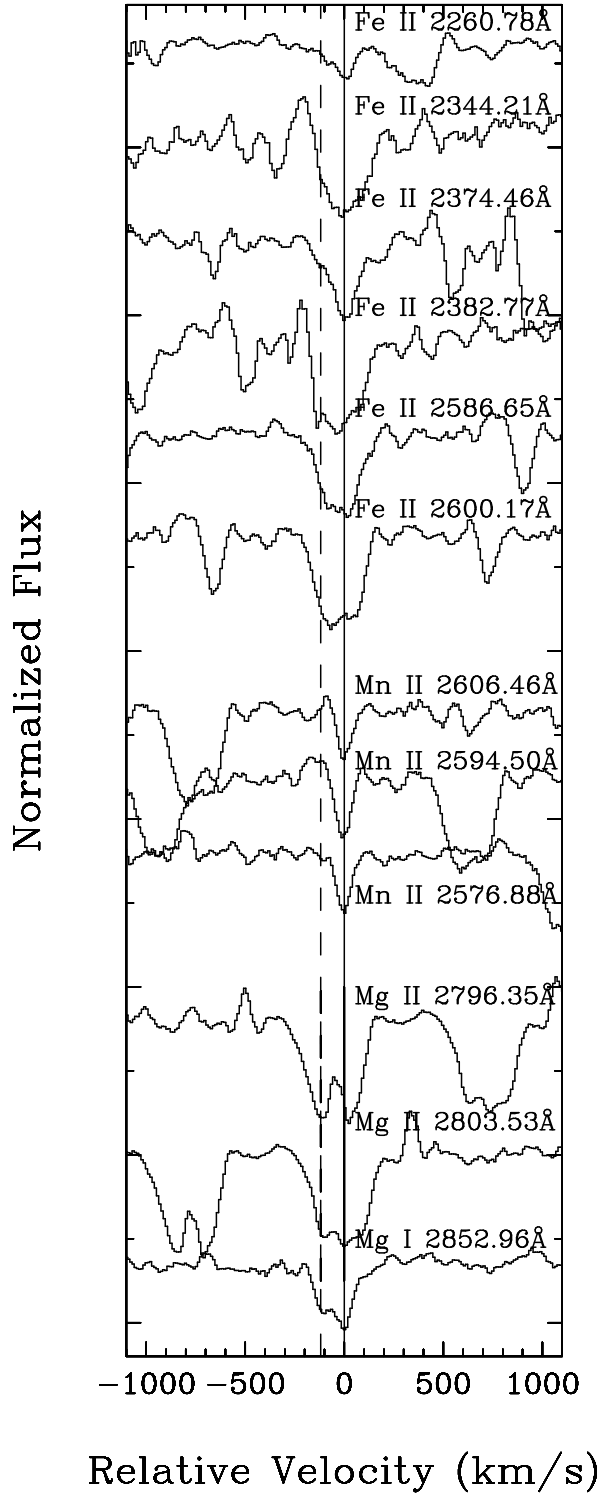}
\caption{Metallic absorption-line profiles in the ESI spectrum plotted in
velocity space, showing evidence of substructure in the most distant 
absorber($z_{1}$). 
As zero velocity we use the systemic redshift, $z=1.47688$.
The {\it dashed line} indicates the average offset of
$v = -119$ km~s$^{-1}$ from the systemic velocity.}
\end{center}
\end{figure}
\clearpage

\begin{figure}[tbp] \label{evol} \figurenum{6}
\begin{center}
\epsscale{.60}
\plotone{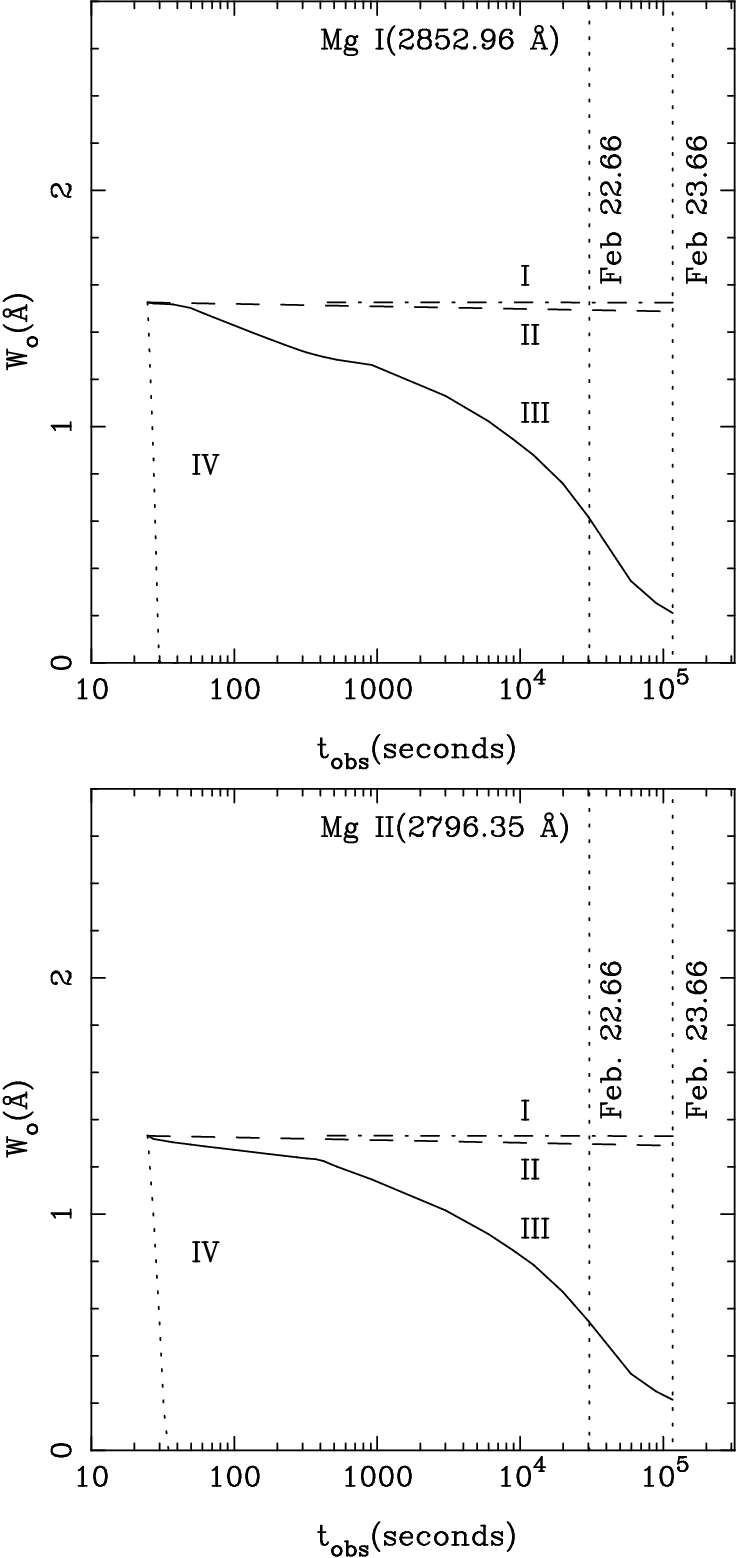}
\caption{Modelled time evolution of the
Mg I (2852.96 \AA) and Mg II (2796.35 \AA)
equivalent widths for different sets of 
initial conditions (Table 5), constrained by
($N_{\rm HI} = 10^{20.44}$ cm$^{-2}$).}
\end{center}
\end{figure}
\clearpage

\begin{figure}[tbp] \label{timeevol} \figurenum{7}
\begin{center}
\epsscale{.60}
\plotone{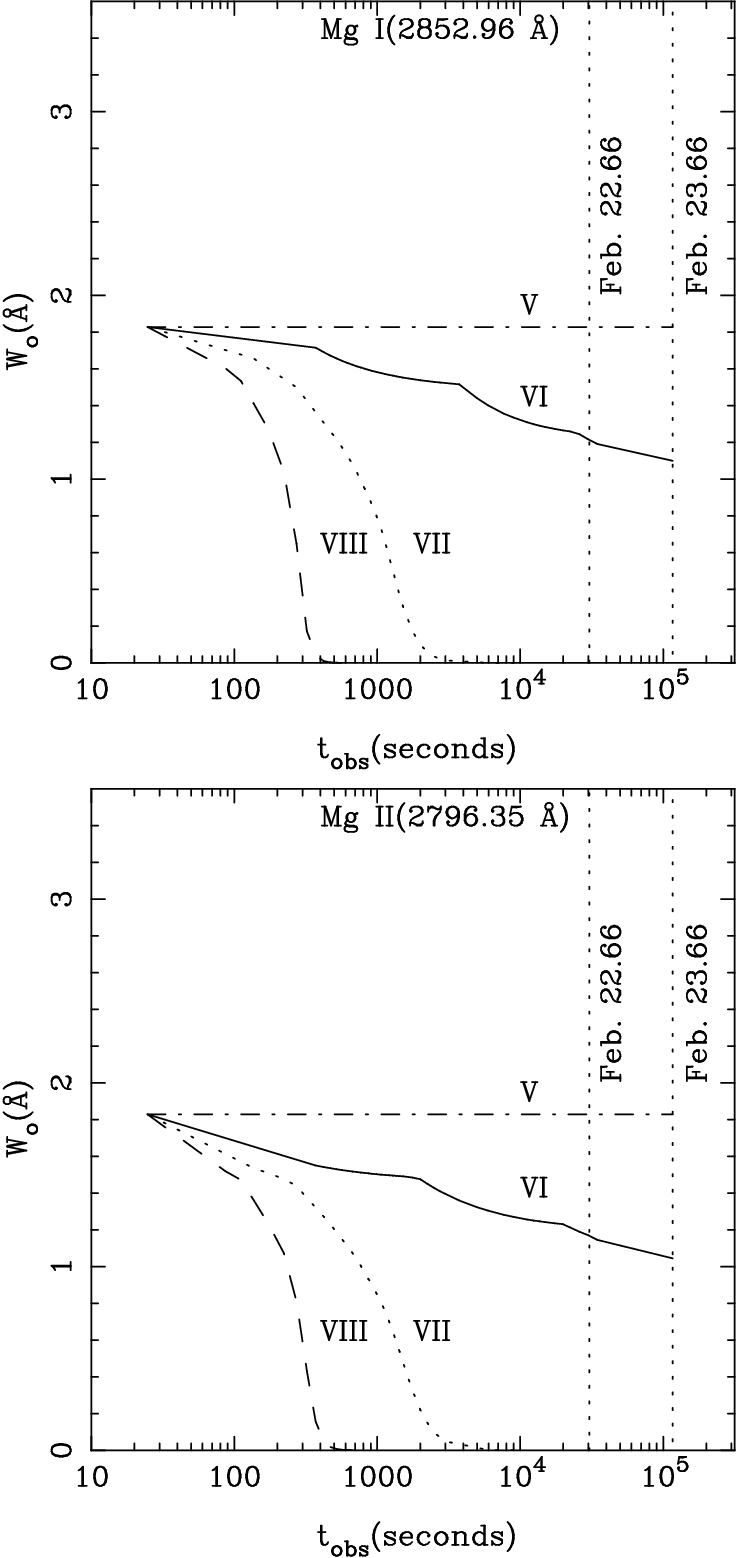}
\caption{Modelled time evolution of the
Mg I (2852.96 \AA) and Mg I(2796.35 \AA)
equivalent widths for different sets of 
initial conditions (Table 5), constrained by 
($N_{\rm HI} = 10^{22}$ cm$^{-2}$).} 
\end{center}
\end{figure}
\clearpage


\clearpage
\begin{references}

\reference{} Aguirre, A. 1999, ApJ, 525, 583

\reference{} Aguirre, A. \& Haiman, Z. 1999, ApJ, 532, 28

\reference{} Anders, E., \& Grevesse, N. 1989 Geochim. Cosmochim. Acta
53, 197

\reference{} Berger, E., Sari, R., Frail, D., \& Kulkarni, S. 2001, in 
Gamma-Ray Bursts in the Afterglow Era (Berlin: Springer), 218

\reference{} Bloom, J. S., Sigurdsson, S., Wijers, R. A. M. J., 
Almaini, O., Tanvir, N. R., \& Johnson, R. A. 1997, MNRAS, 292, 55

\reference{} Bloom, J. S., Djorgovski, S. G., Kulkarni, S. R., \&
Frail, D. A. 1998, ApJ, 507, L25

\reference{} Bloom, J. S., Kulkarni, S. R., \& Djorgovski, S. G. 2002, AJ,
123, 1111 

\reference{} Bloom, J. S., Djorgovski S. G., Halpern J. P., 
Kulkarni S. R., Galama, T. J., Price, P. A., Castro, S. M.~2001, GCN Circular 989

\reference{} B\"ottcher, M., Dermer, C. D., Crider, A. W., \& Liang, E. P. 1999,
A\&A, 343, 111

\reference{} Calzetti, D., Armus, L., Bohlin, R. C., Kinney, A. L., 
Koornneef, J., \& Storchi-Bergmann, T. 2000, ApJ, 533, 682

\reference{} Cardelli, J. A., Clayton, G. C., \& Mathis, J. S. 1989, ApJ, 345,
245

\reference{} Castro, S. et al. 2001a, GCN Circular 999

\reference{} Castro, S., Galama, T. J., Harrison, F. A., Holtzman, J. A., Bloom,
J. S., Djorgovski, S. G., \& Kulkarni S. R. 2001b, astro-ph/0110566

\reference{} Chiappini, C., Matteucci, F. \& Romano, D. 2001, ApJ, 554, 1044

\reference{} Churchill, C. W., Rigby, J. R., Charlton, J. C., 
\& Vogt, S. S. 1999, ApJS, 120, 51

\reference{} Cotera, A. S., Erickson, E. F., Colgan, S. W. J., Simpson, J. P., 
Allen, D. A., \& Burton, M. G. 1996, ApJ, 461, 750

\reference{} Djorgovski, S. G., Kulkarni, S. R., Bloom, J. S., Goodrich, R.,
 Frail, D. A.,  Piro, L., \& Palazzi, E. 1998, ApJ, 508, 17

\reference{} Draine, B. T. 2000, ApJ, 532, 273

\reference{} Figer, D. F., Sungsoo, S. K., Morris, M., Serabyn, E., Rich, R. M.,
\& McLean, I. S. 1999, ApJ, 525, 750

\reference{} Filippenko, A. 1982, PASP, 94, 715

\reference{} Frail, D. A. et al. 2001a, ApJ, 562, L55 

\reference{} Frail, D. A. et al. 2001b, astro-ph/0108436 

\reference{} Fruchter, A. S., Krolik, J. H., \& Rhoads, J. E. 2001, 
ApJ, 563, 597

\reference{} Fryer, C. L., Woosley, S. E., Herant, M., \&
Davies, M. B. 1999, ApJ, 520, 650

\reference{} Galama, T. J., \& Wijers, R. A. M.J. 2001, ApJ, 549, 209

\reference{} Galama, T. J. et al 2002, in preparation

\reference{} Guo, Z., Burrows, D. N., Sanders, W. T., Snowden, S. L., 
\& Penprase, B. E. 1995, ApJ, 453, 256

\reference{} Harrison, F. A., et al. 2001, ApJ, 559, 123

\reference{} Henden, A. 2001a, GCN Circular 961

\reference{} Henden, A. 2001b, GCN Circular 962

\reference{} in 't Zand, J. J. M. et al. 2001, ApJ, 559, 710

\reference{} Janka, H.-T., Eberl, T., Ruffert, M.
\& Fryer, C. L. 1999, ApJ, 527, L39

\reference{} Jenkins, E. B. 1987, in Interstellar Processes 
(Dordrecht: Reidel), 533

\reference{} Jha, S., et al. 2001, ApJ, 554, L155

\reference{} Klebesadel, R. W., Strong, I. B., \& Olson, R. A. 1973, ApJ, 182, 
85

\reference{} Lang, C. C., Goss, W. M., \& Rodriguez, L. F. 2001,ApJ,551,L143

\reference{} Lee, B. C., et al. 2001, ApJ, 561, L183

\reference{} MacFadyen, A. I. \& Woosley, S. E. 1999, ApJ, 524, 262

\reference{} Masetti, N., et al. 2001, A\&A, 374, 382

\reference{} M\'esz\'aros, P., \& Rees, M.J. 1997, ApJ, 476, 232

\reference{} Morton, D. C., \& Bhavsar, S. P. 1979, ApJ, 228, 147

\reference{} Morton, D. C. 1991, ApJS, 77, 119

\reference{} Murakami, H., Koyama, K. \& Maeda, Y. 2001, ApJ, 558, 687


\reference{} Nakamura, T., Umeda, H., Nomoto, K., Thielemann, F.,
\& Burrows, A., et al. 1999, ApJ, 517, 193

\reference{} Nagata, T., Woodward, C. E., Shure, M., \& Kobayashi, N. 1995,
AJ, 109, 1676

\reference{} Panaitescu, A., \& Kumar, P. 2001, ApJ, 560, L49

\reference{} Paczy\'nski, B. 1986, ApJ, 308, L43

\reference{} Perna, R., \& Loeb, A. 1998, ApJ, 501, 467

\reference{} Pettini, M., Ellison, S. L., Steidel, C. C.,
Shapley, A. E., \& Bowen, D. V. 2000, ApJ, 532, 65

\reference{} Piran, T., 1999, Phys. Rep., 314, 575

\reference{} Piro, L., et al. 2000, Science, 290, 955

\reference{} Piro, L., 2001, GCN Circular 959

\reference{} Reichart, D. E. \& Price, P. A. 2002, ApJ, 565, 174

\reference{} Roth, K. C., \& Blades, J. C. 1995, ApJ, L95 

\reference{} Salamanca, I., et al. 2001, astro-ph/0112066

\reference{} Samland, M. 1998, ApJ, 496, 155

\reference{} Sari, R., \& Piran, T. 1997, ApJ, 485, 270

\reference{} Sari, R., \& Esin, A. A. 2001, ApJ, 548, 787

\reference{} Savage, B. D., \& Mathis, J. S. 1979, ARA\&A, 17, 73

\reference{} Savage, B. D., \& Sembach, K. R. 1996, ARA\&A, 34, 279

\reference{} Savaglio, S., Fall, S. M., \& Fiore, F. 2002, 
astro-ph/0203154

\reference{} Scalo, J. \& Wheeler, J. C. 2001, ApJ, 562, 664

\reference{} Schlegel, D. J., Finkbeiner, D. P., \& Davis, M. 1998, ApJ, 500,
525

\reference{} Serabyn, E., Shupe, D., \& Figer, D. F. 1998, Nature, 394, 448

\reference{} Spitzer, L. 1978, in Physical Processes in the Interstellar
Medium (New York: Wiley), 51

\reference{} Stanek, K. Z., et al. 2001, ApJ, 563, 592

\reference{} Steidel, C. C., \& Sargent, W. L. W. 1992, ApJS, 80, 1

\reference{} Timmes, F. X., Woosley, S. E., \& Weaver, T. A. 1995, ApJS, 98,
617

\reference{} van Steenberg, M. E., \& Shull, J. M. 1988, ApJ, 330, 942

\reference{} Verner, D. A., Barthel, P. D., Tytler, D. 1994, A \& A,
108, 287

\reference{} Verner, D. A., Ferland,G. J., Korista, K.T., \& Yakovlev, 
D.G. 1996, ApJ, 465, 487

\reference{} Vietri, M. \& Stella, L. 1999, ApJ, 527, L43

\reference{} Vreeswijk, P. M. et al. 2001, ApJ, 546, 672

\reference{} Waxman, E., \& Draine, B.T. 2000,ApJ, 537,796

\reference{} Welty, D. E., Frisch, P. C., Sonneborn, G. \& York, D. G. 1999,
ApJ, 512, 636

\reference{} Witt, A. N., Thronson, H. A., \& Capuano, J. M. 1992, ApJ, 393, 
611

\reference{} Woosley, S. E. 1993, ApJ, 405, 273




\end{references}
\end{document}